\documentclass{emulateapj} 
\usepackage{graphicx}
\usepackage{multirow}
\usepackage{color}
\usepackage{soul}
\usepackage{ot1patch}
\usepackage{color}
\usepackage[dvipsnames]{xcolor}
\usepackage{verbatim}
\usepackage{ulem}
\usepackage[]{amsmath}

\newcommand{\kyr}{\ensuremath{\,\mathrm{kyr}}}
\newcommand{\myr}{\ensuremath{\,\mathrm{Myr}}}
\newcommand{\gyr}{\ensuremath{\,\mathrm{Gyr}}}

\newcommand{\tage}{\ensuremath{t_{\rm age}}}

\newcommand{\msun}{\ensuremath{\,M_\odot}}

\newcommand{\mns}{\ensuremath{M_{\rm NS}}}
\newcommand{\ma}{\ensuremath{M_{\rm a}}}
\newcommand{\mb}{\ensuremath{M_{\rm b}}}
\newcommand{\mzamsa}{\ensuremath{M_{\rm ZAMS,a}}}
\newcommand{\mzamsb}{\ensuremath{M_{\rm ZAMS,b}}}
\newcommand{\azams}{\ensuremath{a_{\rm ZAMS}}}
\newcommand{\mbhns}{\ensuremath{\,M_{\rm BH/NS}}}
\newcommand{\mmweg}{\ensuremath{M_{\rm MWEG}}}

\newcommand{\mdot}{\ensuremath{\,\dot{M}}}
\newcommand{\macc}{\ensuremath{\mdot_{\rm acc}}}

\newcommand{\medd}{\ensuremath{\mdot_{\rm Edd}}}
\newcommand{\mrlof}{\ensuremath{\mdot_{\rm RLOF}}}
\newcommand{\mt}{\ensuremath{\,\dot{m}}}
\newcommand{\racc}{\ensuremath{\,R_{\rm acc}}}

\newcommand{\rg}{\ensuremath{\,R_{\rm G}}}

\newcommand{\rsun}{\ensuremath{\,R_\odot}}
\newcommand{\msy}{\ensuremath{\msun\mathrm{\; yr}^{-1}}}

\newcommand{\ergs}{\ensuremath{\,\mathrm{erg}\,\mathrm{s}^{-1}}}

\newcommand{\lx}{\ensuremath{L_{\rm X}}}

\newcommand{\ledd}{\ensuremath{L_{\rm Edd}}}
\newcommand{\lxmax}{\ensuremath{L_{\rm X, max}}}
\newcommand{\lxmin}{\ensuremath{L_{\rm X, min}}}

\newcommand{\lxtot}{\ensuremath{L_{\rm X,tot}}}
\newcommand{\lulx}{\ensuremath{10^{39}\ergs}}
\newcommand{\lhlx}{\ensuremath{10^{41}\ergs}}
\newcommand{\zsun}{\ensuremath{Z_\odot}}

\newcommand{\mpc}{\ensuremath{\,\mathrm{Mpc}}}

\newcommand{\startrack}{{\tt StarTrack}}

\newcommand{\dash}{\,\hbox{--}\,}
\newcommand{\nergs}[2]{\ensuremath{#1\times10^{#2}\ergs}}

\newcommand{\nulx}{\#ULX}
\newcommand{\nbhulx}{\#BHULX}
\newcommand{\nnsulx}{\#NSULX}
\newcommand{\ttt}[2]{\ensuremath{#1\times10^{#2}}}
\newcommand{\rnshg}{\ensuremath{\mathcal{R}_{\rm NS,HG}}}
\newcommand{\rnsrg}{\ensuremath{\mathcal{R}_{\rm NS,RG}}}

\newcommand{\rnshehg}{\ensuremath{\mathcal{R}_{\rm NS,evHeS}}}

\newcommand{\rnsms}{\ensuremath{\mathcal{R}_{\rm NS,MS}}}
\newcommand{\rbhms}{\ensuremath{\mathcal{R}_{\rm BH,MS}}}
\newcommand{\rbhhg}{\ensuremath{\mathcal{R}_{\rm BH,HG}}}
\newcommand{\few}[1]{\mbox{{\it a few} $\times\, #1$}}
\newcommand{\several}[1]{\mbox{{\it several} $\times\,#1$}}

\newcommand{\pobsb}{\ensuremath{P_{\rm obs,b}}}

\begin{document}

\title{The origin of the Ultraluminous X-ray Sources}

\author{Grzegorz Wiktorowicz\altaffilmark{1,2},  
        Ma\l{}gorzata Sobolewska\altaffilmark{3}, 
        Jean-Pierre Lasota\altaffilmark{2,4,5},
        Krzysztof Belczynski\altaffilmark{1,2,4}}

 \affil{
     $^{1}$ Astronomical Observatory, Warsaw University, Al.
            Ujazdowskie 4, 00-478 Warsaw, Poland (gwiktoro@astrouw.edu.pl)\\
     $^{2}$ Kavli Institute for Theoretical Physics, Kohn Hall, University of California, Santa Barbara, CA 93106, USA \\
     $^{3}$ Harvard-Smithsonian Center for Astrophysics, 60 Garden Street, Cambridge,
            MA 02138, USA\\
     $^{4}$ Nicolaus Copernicus Astronomical Center, Polish Academy of Sciences, Bartycka 18, 00-716
            Warsaw, Poland\\  
     $^{5}$ Institut d'Astrophysique de Paris, CNRS et Sorbonne Universit\'es, UPMC Paris~06, 
            UMR 7095, 98bis Bd Arago, 75014 Paris, France}
  
\begin{abstract}
Recently, several ultraluminous X-ray (ULX) sources were shown to host a
neutron star (NS) accretor. We perform a suite of evolutionary calculations
which show that, in fact, NSs are the dominant type of ULX accretor.  Although
black holes (BH) dominate early epochs after the star-formation burst, NSs
outweigh them after a few 100 Myr and may appear as late as a few Gyr after the
end of the star formation episode. If star formation is a prolonged and
continuous event (i.e., not a relatively short burst), NS accretors dominate
ULX population at any time in solar metallicity environment, whereas BH
accretors dominate when the metallicity is sub-solar. Our results show a very
clear (and testable) relation between the companion/donor evolutionary stage
and the age of the system. A typical NS ULX consists of a $\sim1.3\,M_\odot$ NS
and $\sim1.0\,M_\odot$ Red Giant. A typical BH ULX consist of a
$\sim8\,M_\odot$ BH and $\sim6\,M_\odot$ main-sequence star. Additionally, we
find that the very luminous ULXs ($L_X\gtrsim10^{41}$ erg/s) are predominantly
BH systems ($\sim9\,M_\odot$) with Hertzsprung gap donors ($\sim2\,M_\odot$).
Nevertheless, some NS ULX systems may also reach extremely high X-ray
luminosities ($\gtrsim10^{41}$ erg/s). 

\end{abstract}


\section{Introduction}

Ultraluminous X-ray source \citep[ULX; for review see][]{Feng1111} is defined by
two  observational properties:
\begin{enumerate}
    \item it is a point-like (i.e. not extended) off-nuclear X-ray source with
        a peak emission localized in the X-ray band;
    \item it emits isotropic equivalent X-ray luminosity in excess of
        $10^{39}\ergs$, which is approximately the Eddington limit (EL) for a
        spherically accreting stellar-mass black hole (sMBH;
        $\sim10\msun$).
\end{enumerate}

Until recently the main problem with unveiling the nature of the ULXs stemmed
from the absence of reliable measurements of the masses of the accreting
compact objects.  This has changed with the discovery of pulsing ultraluminous
X-ray sources (PULXs) which contain neutron stars (NS). The first X-ray pulsar
was discovered in M82 X-2 \citep{Bachetti1410}. Then, two other PULXs have been
identified: P13 in NGC 7793 \citep{Furst1611,Israel1703}, and NGC5907 ULX1,
\citep{Israel1609,Furst1701}.  All these PULXs show regular pulses with periods
$\sim 1$\,s, characteristic of NS accretors.

These discoveries proved that some ULXs do contain stellar-mass compact
objects. Moreover, it is possible that NSs may reside in a vast majority of the
ULXs \citep[NSULXs;][]{Kluzniak1503,King1605,King1702}.  Existence of PULXs
provides strong evidence in support of the models involving a super-Eddington
accretion onto a compact star
\citep[e.g.,][]{Begelman0204,Poutanen0705,King0902} as an explanation of the
ULXs phenomenon \citep[e.g.][]{Kording0201}.

However, apparent super-Eddington luminosities can also be reached without
breaching the EL if the radiation of an accreting compact star is beamed
\citep[e.g.,][]{King0105,Poutanen0705}.  Detailed accretion disk simulations
appear to support the importance of beaming
\citep[e.g.,][]{Ohsuga0507,Ohsuga1107,Jiang1412,Sadowski1511}. 

In this study we perform massive simulations in order to uncover the nature of
the stars forming the ULX population. Motivated by the recent observational
progress, we limited our investigation to the stellar-mass compact objects with
super-Eddington and/or beamed emission.

\section{Simulations}\label{sec:simulations}

The calculations were performed with the use of \startrack\ population
synthesis code \citep{Belczynski0206b,Belczynski0801} with significant updates
described in \citet{Dominik1211,Wiktorowicz1409}.

We start the evolution of each system from the Zero Age Main Sequence (ZAMS),
which we assume to happen at the same moment for both stars. The primary has an
initial mass of $M_a=6\dash150\msun$ drawn from a power-law distribution with
index $-2.7$ \citep{Kroupa0312}. The mass of the secondary ($M_b$) is chosen
from the $0.08\dash150\msun$ range to preserve a uniform mass-ratio
distribution. Typical predecessors of X-ray binaries (XRB) with NS or BH
accretors should reside in such a range of masses. The initial separations are
uniform in logarithm \citep[$P(a)\sim1/a$;][]{Abt83}. The eccentricities'
distribution is the thermal-equilibrium \citep{Duquennoy9108}. The binary
fraction was set to $50\%$ for stars with ZAMS mass below $10\msun$ and $100\%$
for systems with more massive primaries. We are aware of the results of
\citet{Sana1207} who provide other relations for initial parameters
distributions, but their results were obtained for stars with masses
$15\dash60\msun$ and were limited to solar metallicity. Therefore, we decided
to keep the previously established  distributions. For comparison between our
adopted and \citet{Sana1207} initial distributions the reader may refer to
\citet{deMink1511}. 

We simulated the evolution of $2\times 10^7$ binary systems for every model and
scaled the results to the Milky-Way equivalent galaxy
($\mmweg=\ttt{6}{10}\msun$ \citep{Licquia1506}; star formation rate (SFR) equal
$6.0\msy$ for constant star formation (SF) during $10\gyr$, and $600\msy$ for
burst SF with duration of $100\myr$). We do our study for solar metallicity
(\zsun), $10\%$ of solar (\zsun/10), and $1\%$ of solar (\zsun/100).

\subsection{Accretion model}\label{sec:acc_model}

We implemented a model based on nonlinear scaling of X-ray luminosity and mass
accretion rate \citep[e.g.][]{Shakura73}, 

\begin{equation}\label{eq:lxtot}
    \lxtot=\ledd(1+\ln\mt), 
\end{equation}

\noindent where \mt\ is the mass transfer (MT) rate in Eddington rate units.
Mass accretion onto compact object (\macc) is limited to \medd\ and the rest of
mass transferred to the system is lost in a wind from the inner disk region.
Such outflows from ULXs were recently observed by \citet{Pinto1605}. 

\subsection{Beaming model}\label{sec:beaming_model}

\begin{figure}[t]
    \centering
    \rule{\columnwidth}{1pt}
    \includegraphics[width=0.9\columnwidth]{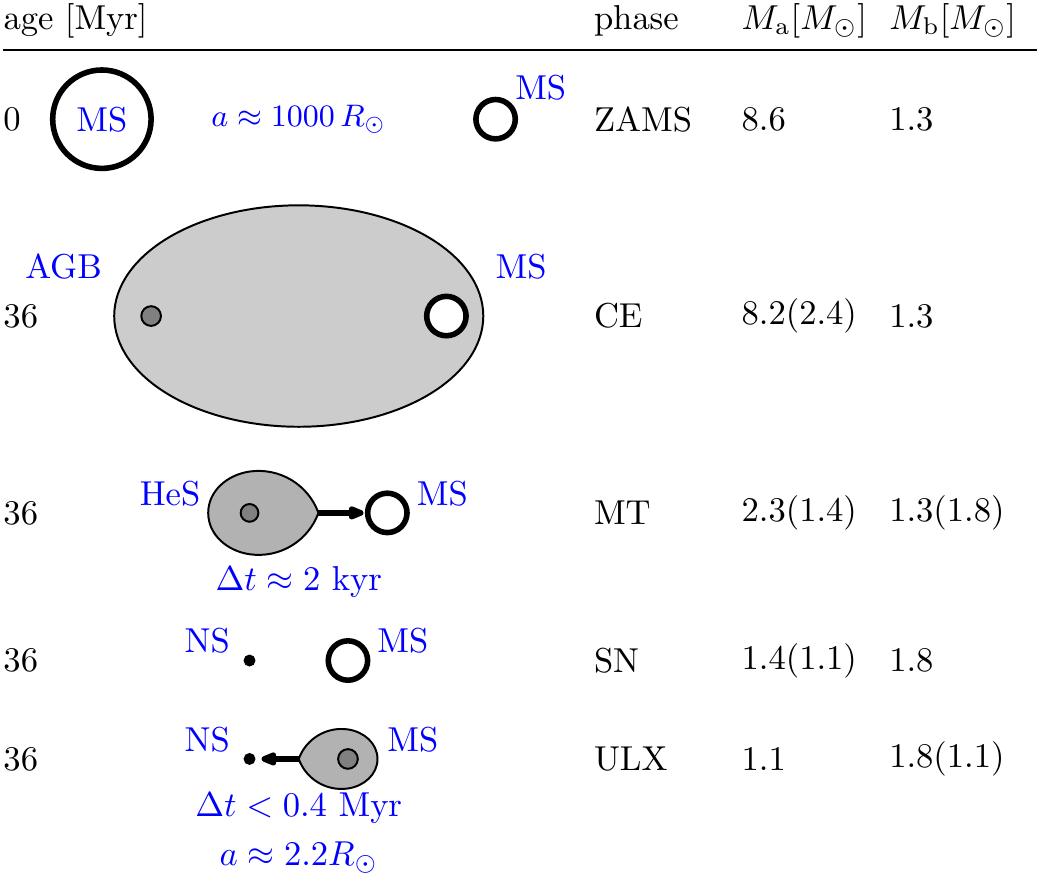}
    \rule{\columnwidth}{1pt}
    \includegraphics[width=0.9\columnwidth]{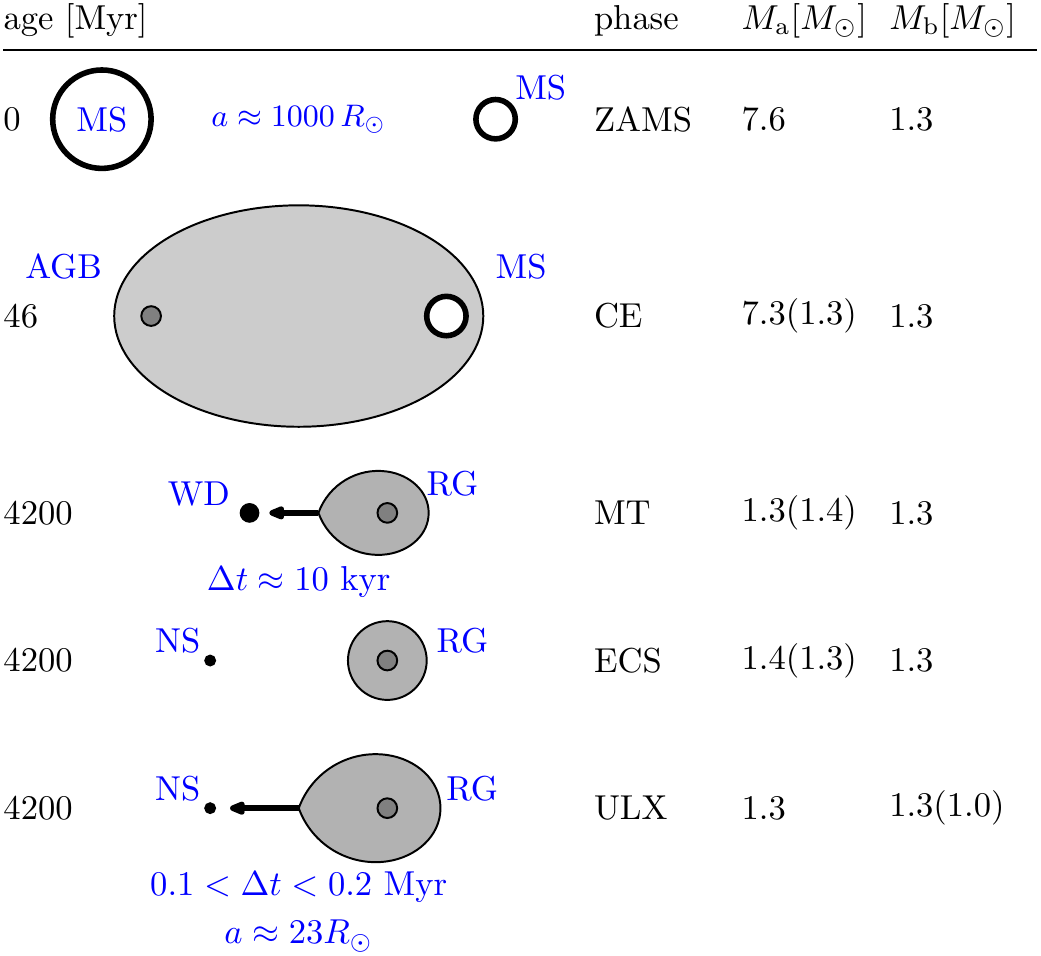}
    \rule{\columnwidth}{1pt}
    \includegraphics[width=0.9\columnwidth]{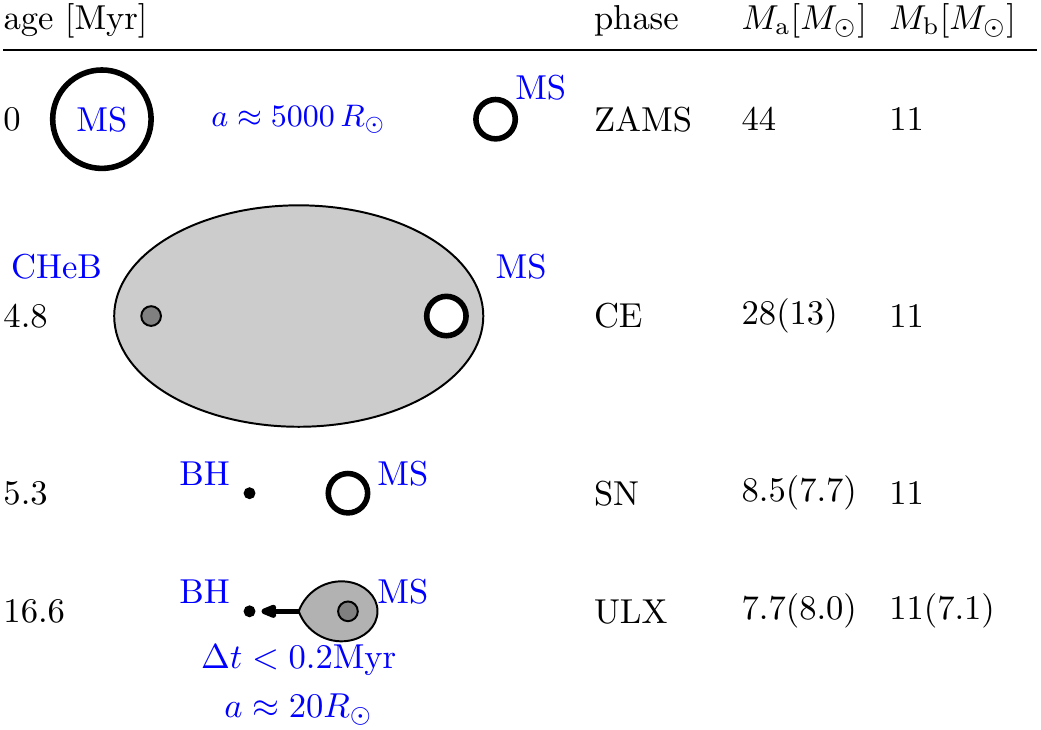}
    \caption{Schematic representation of the dominant NSULX evolutionary routes
    in young (top; \rnsms) and old (middle; \rnsrg) stellar populations.
    Bottom: typical BHULX evolution (SF regions). The age and masses values
    reflect those in a typical system. For explanation of abbreviations see
    text (Sec.~\ref{sec:nsulx}). For phases in which the mass of one of the
    components changes rapidly two values are provided: the mass at the
    beginning of a phase and at the end (in parethesis).  }
    \label{fig:rulx}
\end{figure}

The collimation of radiation from the innermost parts of the accretion disk,
i.e. beaming, has a prodigious influence on the {\it apparent}\ isotropic
luminosity.  The beaming factor is defined as $b=\Omega / 4\pi$, where $\Omega$
is the solid angle of emission \citep[e.g.][]{King0105}. If we assume that
there are two conical beams with opening angle $\theta$, then
$\Omega=4\pi(1-\cos\theta/2)$. 

The isotropic equivalent X-ray luminosity ($\lx$) can be expressed as

\begin{equation}\label{eq:lxiso}
    \lx=\frac{\lxtot}{b}.
\end{equation}

Under assumption of the isotropic distribution of disk inclinations in space
\citep[see][]{King0902}, the probability of observing a source along the beam
is equal $b$.

The results of theoretical studies as well as the outcomes of detailed modeling
of accretion disks favor the dependence of beaming on the mass transfer rate
\citep[e.g.,][]{Lasota16}.  \citet{King0902} proposed that $b$ scales as

\begin{equation}\label{eq:b}
\begin{aligned}
    &b\sim\frac{73}{\mt^2} & \mt\geq8.5,\\
    &b=1 & \mt<8.5.
\end{aligned}
\end{equation}

For very high MT rates this prescription may give an extremely small value of
$b$, i.e. strong beaming. For example, a Galactic ULX candidate SS433 has
$\mt\approx3000\dash10^4$ \citep[e.g.][]{Fabrika04} for which Eqs.~\ref{eq:b}
and \ref{eq:lxiso} give $b\approx\ttt{7}{-7}\dash\ttt{8}{-6}$ and
$\lx\approx10^{46}\ergs$. Therefore, we will assume saturation at $\mt = 150$.
For higher mass accretion rates we will consider that the beaming is constant
and equal $b\approx\ttt{3.2}{-3}\,(\theta\approx9^\circ)$.  Theoretical
background for beaming saturation was presented by \citet{Lasota1603}.

\section{Results} \label{sec:results}   

\begin{deluxetable*}{lcccccccc}
    \tablewidth{\textwidth}
    \tablecaption{Typical parameters of ULXs}

    \tablehead{ & \multicolumn{5}{c}{Present} & \multicolumn{3}{c}{ZAMS}\\
               Route  &  t[Myr] & $\Delta t$[Myr] & \ma[\msun] & \mb[\msun] & a[\rsun] & \mzamsa[\msun] & \mzamsb[\msun] & \azams[\rsun] }
    \startdata
    \multicolumn{9}{c}{$Z=\zsun$}\\
    \rbhms & $11\dash22$ & $<0.2$     & $7.7\dash8.6$    & $5.9\dash7.2$      & $18\dash22$            & $40\dash50$  & $6.6\dash12$ & $3800\dash4600$\\
     \multicolumn{9}{l}{\hspace{0.2cm}\it ev. route:\rm\ CE1(4-1;7/8-1) SN1 MT2(14-1)}\\
    \rnsms & $6\dash38$ & $\lesssim0.4$ & $1.1\dash1.3$ & $1.1\dash1.4$ & $2.2\dash3.6$ & $8.2\dash9.3$ & $1.3\dash1.7$ & $700\dash1200$\\
            \multicolumn{9}{l}{\hspace{0.2cm}\it ev. route:\rm\ CE1(3/4/5-1;7/8-1) SN1 MT2(13-1)}\\
    \rnshg & $430\dash730$ &$0.3\dash0.8$     & $\sim1.3$    & $0.6\dash1.0$      & $7.3\dash17$            & $7.9\dash8.3$  & $1.6\dash2.0$ & $800\dash1200$\\
            \multicolumn{9}{l}{\hspace{0.2cm}\it ev. route:\rm\ CE1(3/4/5-1;7/8-1) SN1 MT2(13-1/2)}\\
    \rnsrg &  $2300\dash4300$ & $0.1\dash0.2$    & $\sim1.3$   & $\sim1.0$ & $35\dash50$            & $7.5\dash7.9$   & $1.2\dash1.5$     & $1100\dash1700$\\
            \multicolumn{9}{l}{\hspace{0.2cm}\it ev. route:\rm\ CE1(6-1;12-1) MT2(12-3) AICNS1 MT2(13-3)}\\
                                        
    \multicolumn{9}{c}{$Z=\zsun/10$}\\
    \rbhms      &  $4\dash37$     & $<27$   & $5.7\dash8.9$    & $5.6\dash7.8$ &
    $14\dash19$            & $26\dash37$  & $6.0\dash11$ & $1800\dash3700$\\
    \rnsms      & $9\dash55$ & $\lesssim0.8$ & $1.1\dash1.3$ & $1.2\dash1.5$ & $1.6\dash2.7$ & $7.2\dash12$ & $1.5\dash 1.9$ & $1000\dash2600$\\
    \rnshg      & $400\dash820$      & $\lesssim0.6$   & $1.2\dash1.3$    & $0.6\dash0.9$  & $9.3\dash22 $         & $6\dash8$  & $1.3\dash1.8$ & $600\dash1500$\\
    \rnsrg      & $1400\dash2600$   & $\lesssim0.1$  & $\sim1.3$   & $\sim1.0$      & $35\dash45$            & $6.3\dash6.8$   & $1.3\dash1.6$     & $1700\dash2700$\\

    \multicolumn{9}{c}{$Z=\zsun/100$}\\
    \rbhms   & $8.7\dash19$    & $<1.0$    & $12\dash21$   & $7\dash11$ &
    $12\dash17$     & $35\dash55$  & $9.1\dash14$ & $540\dash2000$\\
    \rnsms  & $15\dash59$  & $0.6\dash1.3$ & $1.2\dash1.4$ & $0.9\dash1.0$ & $2.8\dash3.5$ & $7.2\dash8.2$ & $1.7\dash2.0$ & $500\dash900$\\
    \rnshg      & $400\dash690$      & $\lesssim0.4$   & $\sim1.3$    & $0.6\dash0.8$  & $17\dash35$      & $6.0\dash6.3$  & $1.7\dash2.1$ & $700\dash900$\\
    \rnsrg      & $1200\dash2100$   & $\lesssim0.1$  & $\sim1.3$   & $\sim1.0$      & $30\dash45$     & $6.0\dash6.2$   & $1.3\dash1.7$     & $900\dash1300$\\

    \enddata
    \tablecomments{Typical present and ZAMS parameters of most common ULX
    evolutionary routes. Ranges represent $50\%$ of values in present day
    populations. The parameter columns represent respectively: Age of the
    system in ULX phase ($t$), duration of ULX phase ($\Delta t$), compact
    object mass (\ma), donor mass (\mb), separation ($a$), primary mass on ZAMS
    (\mzamsa), secondary mass on ZAMS (\mzamsb), separation on ZAMS (\azams).
    The schematic evolutionary routes provided for $Z=\zsun$ are the same for
    other metallicities. Symbols meaning \citep{Belczynski0801}: CE1~-~common
    envelope (donor: primary); MT2~-~mass transfer (donor: secondary);
    SN1~-~supernova (primary); AICNS1~-~accretion induced collapse (primary);
    1~-~MS; 2~-~HG; 3~-~RG; 4~-~CHeB; 5~-~early AGB; 6~-~thermal-pulsing AGB;
    7~-~HeS; 8~-~evHeS; 12~-~ONeWD; 13~-~NS; 14~-~BH.}
\label{tab:params}
\end{deluxetable*}

In this section we provide results for the most typical systems
and focus on the standard model (Sec.~\ref{sec:simulations}). In the Appendix,
we provide additional characteristics of the simulated population
of ULXs and results for other models.

\subsection{Formation of NSULXs}\label{sec:nsulx}

\begin{figure}[h]
    \centering
    \includegraphics[width=0.8\columnwidth]{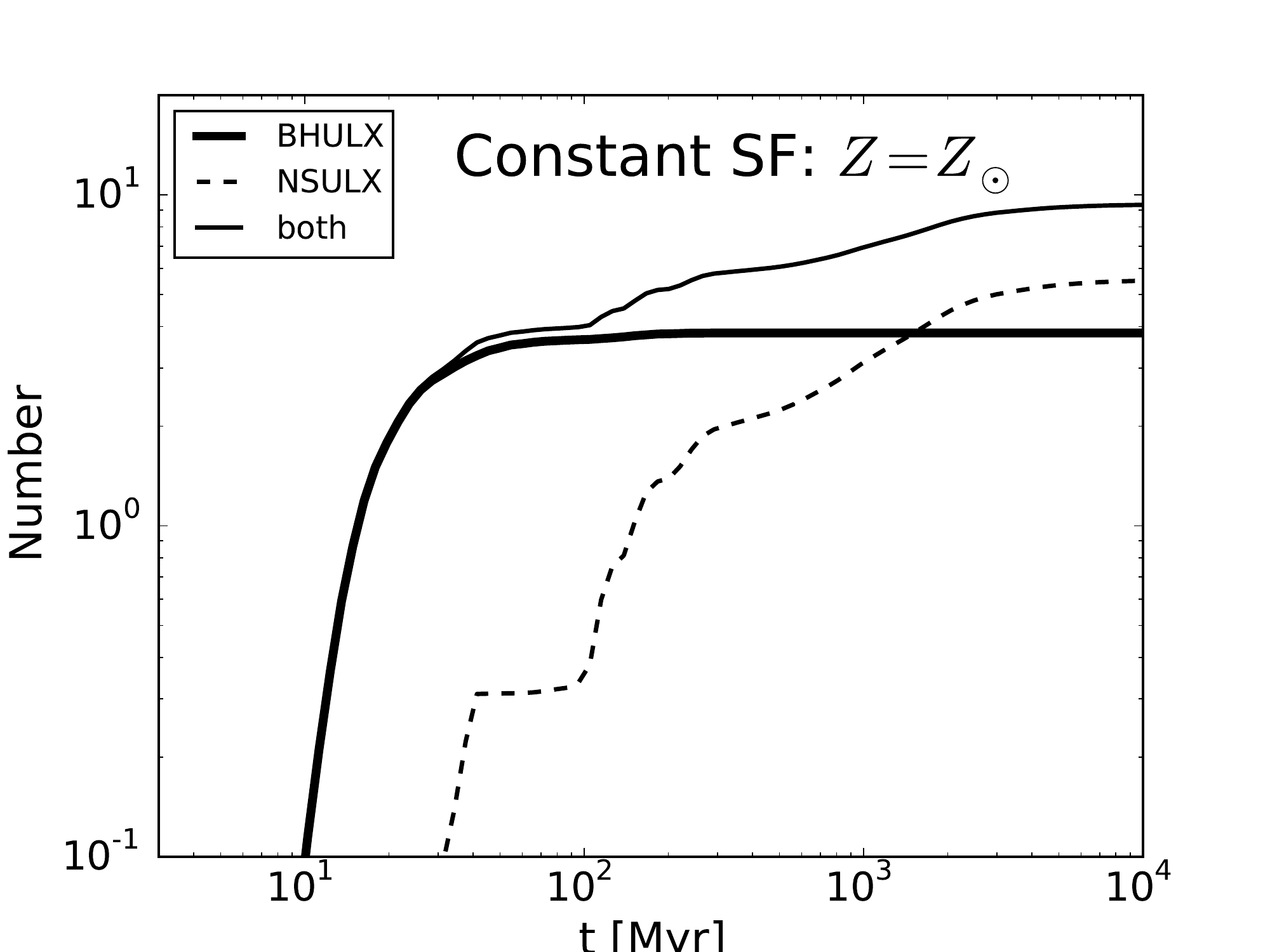}
    \includegraphics[width=0.8\columnwidth]{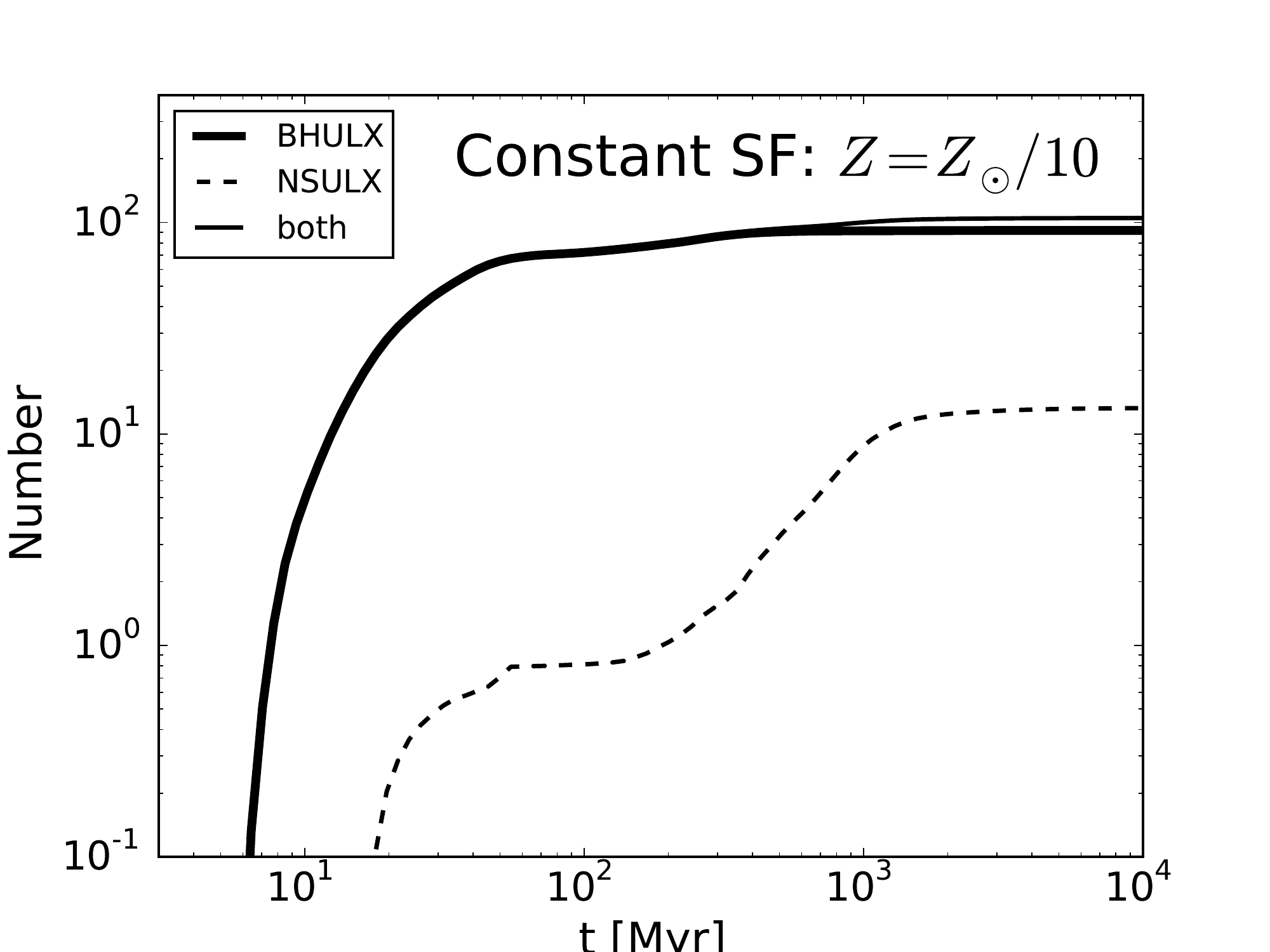}
    \includegraphics[width=0.8\columnwidth]{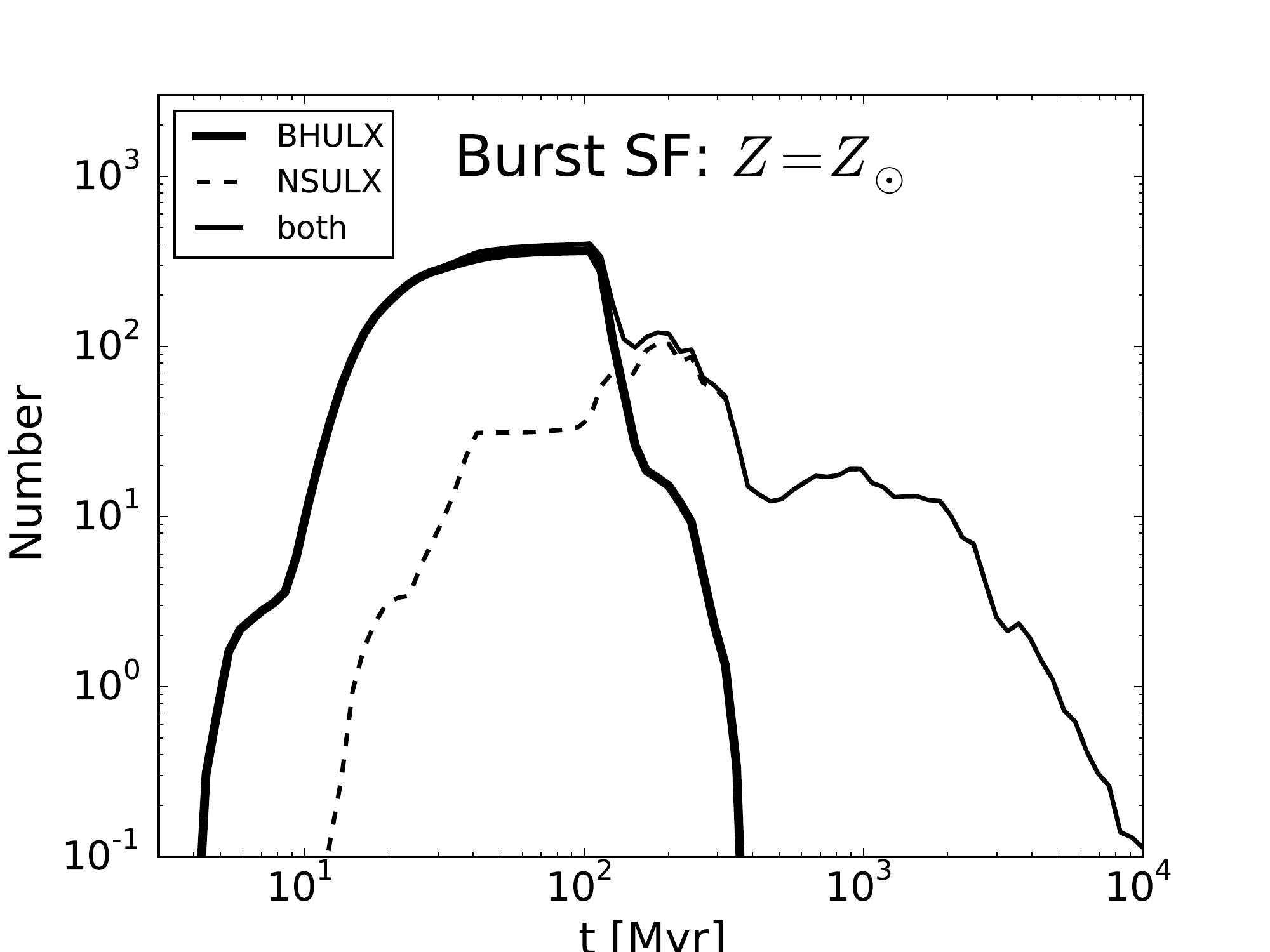}
    \includegraphics[width=0.8\columnwidth]{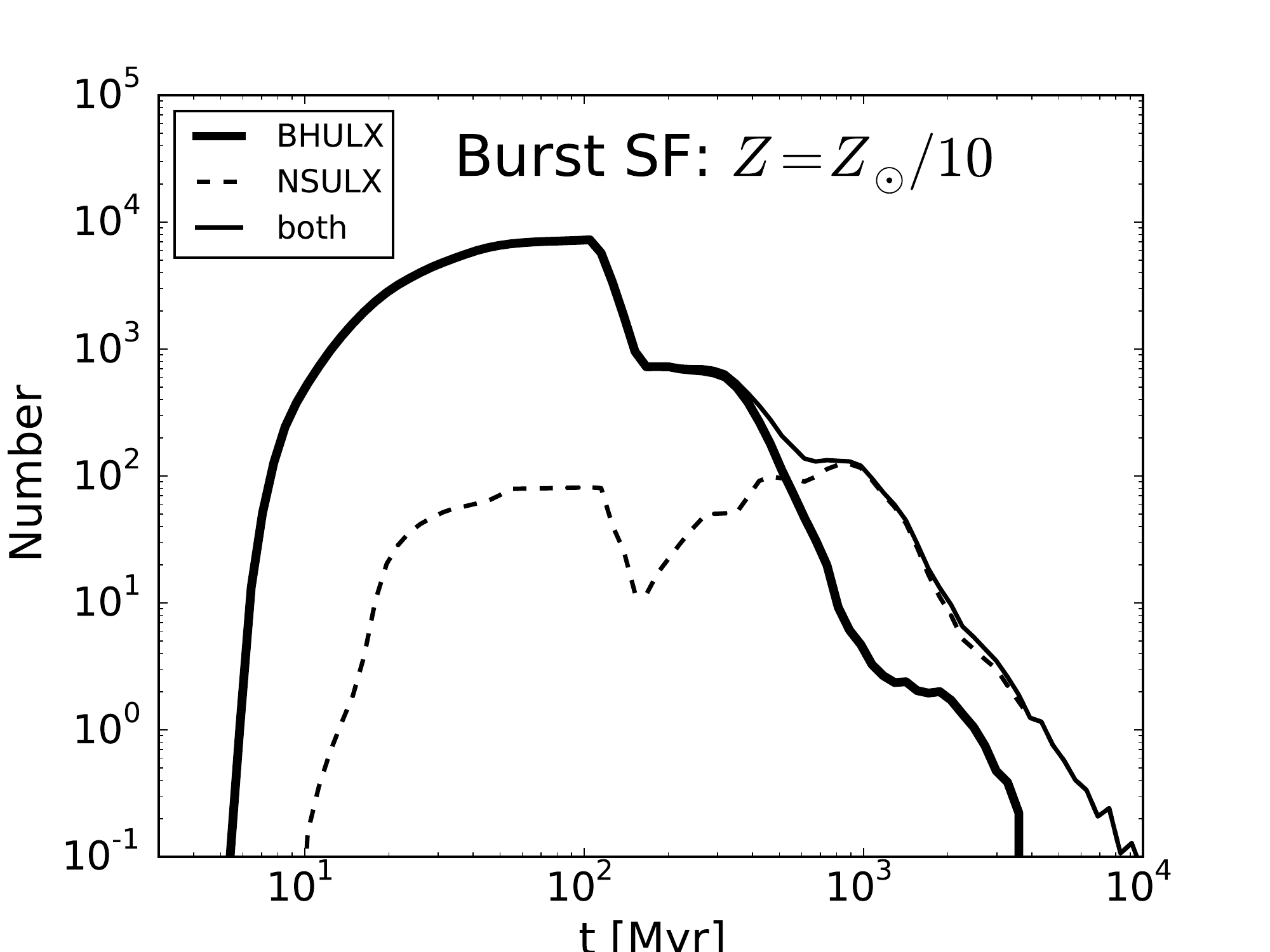}
    \caption{Time evolution of the number of ULXs since the beginning of star
    formation. The BHULXs (thick line) appear early, but $100\dash1000\myr$
    later the ULX population becomes dominated by the NSULXs (dashed line),
    except constant SF models with sub-solar metallicities. Number of ULXs for
    a star forming mass of \ttt{6}{10}\msun (see Sec.~\ref{sec:simulations}).
    Presented values are for all ULXs (including less typical routes, which are
    not presented in Tab.~\ref{tab:params}). For the relative abundance of
    young and old ULXs see Fig.~\ref{fig:num_yo}.}
    \label{fig:nulx}
\end{figure}

According to our results, NSULX are present in ULX populations of all ages and
metallicities.  The first NSULXs form as early as $\sim6\myr$ after the start
of SF, but the ULX phase may occur also in a very old system
($\tage\sim5\gyr$).  Dominant evolutionary routes leading to the occurrence of
high MT rate and formation of a NSULX depend on the age of the system at the
time of the ULX phase.  Below we describe the typical routes for early
(\rnsms), mid-age (\rnshg), and old (\rnsrg) stellar populations (see also
Fig.~\ref{fig:rulx}). Ranges represent $50\%$ of values in present day
populations \footnote{i.e., population as we predict them to look like for an
observer able to see all the objects at the same time. Problems related to
light propagation (e.g. light speed, interstellar absorption) are not discussed
in this study.}.

\paragraph{$\mathbf{6\dash59\myr}$; Route \rnsms; Typical companion MS} On ZAMS
the masses of the primary and secondary are $7.2\dash12\msun$ and
$1.3\dash2.0\msun$, respectively. The initial separation, $500\dash2600\rsun$,
shrinks during the CE phase commenced by the primary. After the CE phase, the
primary forms a NS through a core collapse supernova explosion (SN). The natal
kick leaves the system on an orbit that is tight enough for the secondary to
fill its Roche lobe (RL) within about $2 \kyr$ due to nuclear expansion and
gravitational radiation (GR) before it leaves the main sequence (MS). During
the RLOF the companion is comparable in mass ($0.9\dash1.5\msun$) to the NS
($1.1\dash1.4\msun$), thus the MT is stable and may be sustained as long as
$\sim1.3\myr$ (see Fig.~\ref{fig:rulx}, top). 

\paragraph{$\mathbf{400\dash820\myr}$; Route \rnshg; Typical companion HG} In
approximately $54\dash73\%$ percent of the systems following the \rnsms\ route
the RLOF during the MS phase, if present, is not strong enough to power an ULX.
These secondaries start to reach their terminal-age MS and expand rapidly
several hundred $\myr$ after the start of the SF. If the separation is short
enough, they will fill their RL. Thus, at this time, a Hertzsprung-Gap (HG)
star becomes the most common NSULX companion.

\paragraph{$\mathbf{1200\dash 4300\myr}$; Route \rnsrg; Typical companion RG}
The evolution of the late-time NSULXs is significantly different than that of
the earlier ULXs. After a CE phase the primary forms an Oxygen-Neon White Dwarf
(ONeWD) instead of a NS, and thus the mass of the primary is on average lower
compared with routes \rnsms\ and \rnshg. However, when the secondary becomes a
Red Giant (RG) and fills its RL of $\sim10\dash20\rsun$, the primary accretes
additional mass and forms a NS due to an accretion induced collapse (AIC).
Afterwards, the RG refills its RL and a short ULX phase occurs
(Fig.~\ref{fig:rulx}, middle).

\subsubsection{Formation of BHULXs}\label{sec:formation_bhulx}

In general, BHULXs are a minority in our results. Nevertheless, they dominate
the ULX population during the initial phase of the constant SF (exception being
populations with low ($\leq\zsun/10$) metallicity), and during the SF burst.
They are also more numerous than NSULXs among the ULXs with luminosity in
excess of $10^{41}$\,erg\,s$^{-1}$ (see Secs.~\ref{sec:nulx} and
\ref{sec:hlx}).

\paragraph{$\mathbf{4\dash 37\myr}$; Route \rbhms; Typical companion MS} The
initial binary on ZAMS consist of a massive primary ($26\dash55\msun$) and an
intermediate-mass secondary ($6\dash14\msun$). The initial semi-major axis is
$\sim540\dash4000\rsun$.

The heavier star evolves very quickly and during the CHeB phase fills its RL
commencing the CE episode. As a result, the primary is deprived of its hydrogen
envelope and the separations drops to \few{10\rsun}. A direct collapse occurs
shortly afterwards. The combined action of the companion's nuclear expansion
and gravitational radiation (GR) leads to a RLOF. The phase of ULX emission is
short, $\lesssim1.0\myr$ (Fig.~\ref{fig:rulx}, bottom).

\bigskip

In the case of the \rbhms\ channel it is essential that the heavy primary
expands and fills a RL of \several{100\rsun} what terminates the expansion.
There may be some observational evidence that for the most massive stars ($M >
40\msun$) the large dimensions are avoided \citep{Mennekens1404}. Additionally,
the massive stars are almost all found in close interacting binaries
\citep{Sana1207}, where very quickly their envelope is removed by RLOF and/or
CE. It is naturally expected in our models. Since the post-MS expansion is very
rapid ($\sim200\kyr$), and envelope removal is even quicker ($\sim1\kyr$) we do
not expect many of very massive stars to show up at large radius (as red
supergiants).  Moreover, rapid rotation, which we do not take into account, may
additionally limit expansion of some of the most massive stars.

\subsection{Number of ULXs}\label{sec:nulx}

By the number of ULXs (\nulx) we understand the predicted number of observed
systems at the present time. We find that \nulx\ strongly depends on the SFR
history (Fig.~\ref{fig:nulx}).

Constant SFR naturally leads to a constant growth in \nulx, which starts early
after the start of SF ($\sim5\myr$) and after a few \gyr\ reaches a constant
value of $\nulx=9.3$, $100$, and $56$ for \zsun,\zsun/10, and \zsun/100,
respectively (see Sec.~\ref{sec:disc_Z}).

BHULXs appear first in both constant and burst SF scenarios.  In the former
case, the number of BHULXs increases to the age of a few hundreds $\myr$, at
which time it reaches a constant value of 3.8, 92, 51 for $\zsun$, $\zsun/10$,
and $\zsun/100$, respectively.  In the case of a burst SF, the formation time
of BHULXs is limited nearly exactly to the duration of the burst
($\lesssim200\myr$).

NSULXs resulting from different evolutionary routes described in
Sec.~\ref{sec:nsulx} appear sequentially in their chronological order and form
the characteristic steep sections (constant SF) or humps (burst SF) at
$\sim100\myr$ (\rnsms), $\sim200\myr$ (\rnshg), and $\sim1\gyr$ (\rnsrg).

ULX populations younger than $\sim100\myr$ are dominated by the BHULX.  For
constant SF, the NSULXs become more abundant then the BHULXs ($58\%$ of the ULX
accretors) only if $Z=\zsun$. For lower metallicities, the \nbhulx\ grows
rapidly during the first few hundreds $\myr$ whereas the rate of the NSULXs
formation is much lower, and they are not able to match the abundance of the
BHULXs during the next $10\gyr$. As a result, the NSULXs account for not more
than $\sim12\%$ and $\sim8\%$ of all ULXs for \zsun/10 and \zsun/100,
respectively.  However, for burst SF and all investigated metallicities, the
NSULXs exceed the BHULXs in numbers within $\lesssim1\gyr$, and after a few
\gyr\ they become the only surviving ULXs.

\subsection{ULXs with luminosities exceeding 10$^{41}$ erg
s$^{-1}$}\label{sec:hlx}

Several surveys found a break in the X-ray luminosity function (XLF) at
$\lesssim\ttt{2}{10}\ergs$, which may point toward a different populations
below and above this luminosity
\citep[e.g.,][]{Grimm0303,Swartz1111,Mineo1201}.  Thus, it is interesting to
point out that hyper-luminous X-ray sources (HLX) defined as ULXs with
$\lx\gtrsim\lhlx$, thus clearly above the $\ttt{2}{10}\ergs$ threshold, have
different donors than the ULXs with $\lx\ll\lhlx$.  We find that $52 \dash
84\%$ of the ULXs (described in Sec. \ref{sec:results}) have luminosities in
the range $(1 \dash 3)\times 10^{39}$\,erg\,s$^{-1}$.

Approximately $90\%$ of HLXs are BHULXs with HG donors. NS accretors in HLXs
may obtain nearly as high luminosities as BH accretors and are typically
accompanied by evolved helium stars (evHeS).  Noteworthy, the evolutionary HLX
routes match those reported for HLXs with $\lx\geq10^{42}\ergs$ in
\citet{Wiktorowicz1509}. The current results, however, are based on a much
wider range of models and more physical treatment of NS/BH accretion, and we
include here the HLX evolutionary paths for completeness (Tab.~\ref{tab:params_hlx}).  Notice that the PULX
NGC5907 ULX1 has a luminosity $\gtrsim 10^{41}$ erg s$^{-1}$\citep{Furst1611}.

\begin{deluxetable*}{lcccccccc}
    \tablewidth{\textwidth}
    \tablecaption{Typical parameters of HLXs}

    \tablehead{ & \multicolumn{5}{c}{Present} & \multicolumn{3}{c}{ZAMS}\\
               Route  &  t[\myr] & $\Delta t$[\myr] & \ma[\msun] & \mb[\msun] & a[\rsun] & \mzamsa[\msun] & \mzamsb[\msun] & \azams[\rsun] }
    \startdata
    \multicolumn{9}{c}{$Z=\zsun$}\\
    \rbhhg      & $15\dash40$   & $\lesssim 0.08$    & $8.0\dash10$    & $1.7\dash3.3$      & $20\dash90$            & $40\dash50$  & $3.8\dash8.2$ & $3700\dash4600$\\
       \multicolumn{9}{l}{\hspace{0.2cm}\it ev. route:\rm\ CE1(4/5-1;7/8-1) SN1 MT2(14-1/2)}\\
    \rnshehg    & $17\dash40$ & $\lesssim 0.001$    & $1.3\dash1.4$ &  $1.7\dash2.6$ &$\lesssim4.4$ & $10\dash11$ & $8.5\dash10$ & $30\dash300$\\
       \multicolumn{9}{l}{\hspace{0.2cm}\it ev. route:\rm\ MT1(2/3/4/5/6-1/2/4) SN1 CE2(13-3/4/5;13-7/8) MT2(13-8/9)}\\
                                        
    \multicolumn{9}{c}{$Z=\zsun/10$}\\
    \rbhhg   &  $20\dash60$  & $\lesssim 0.06$    & $7.2\dash9.5$    & $1.7\dash3.3$ & $12\dash120$  & $25\dash35$  & $4.0\dash7.0$ & $2000\dash5000$\\
    \rnshehg    & $40\dash60$ & $\lesssim 0.001$    & $\sim1.3$ &  $1.5\dash2.0$ &$8.4\dash13$ & $8.8\dash9.5$ & $5.5\dash7.3$ & $170\dash430$\\
    \multicolumn{9}{c}{$Z=\zsun/100$}\\
    \rbhhg   & $20\dash45$    & $\lesssim 0.04$    & $10\dash18$   & $1.2\dash3.7$ & $10\dash75$     & $25\dash45$  & $5.0\dash9.0$ & $700\dash2600$\\
    \rnshehg    & $50\dash75$ & $\lesssim 0.001$    & $\sim1.3$ &  $1.9\dash2.5$ &$2.1\dash6.9$ & $6.4\dash6.8$ & $5.5\dash6.2$ & $170\dash280$\\
    \enddata
    \tablecomments{Typical present and ZAMS parameters of typical most luminous
        ULXs (HLX; routes
        \rbhhg\ and \rnshehg). The table is organized in the same way as
        Tab.~\ref{tab:params}.}
\label{tab:params_hlx}
\end{deluxetable*}

\begin{figure}[h]
    \centering
    \rule{\columnwidth}{1pt}
    \includegraphics[width=0.9\columnwidth]{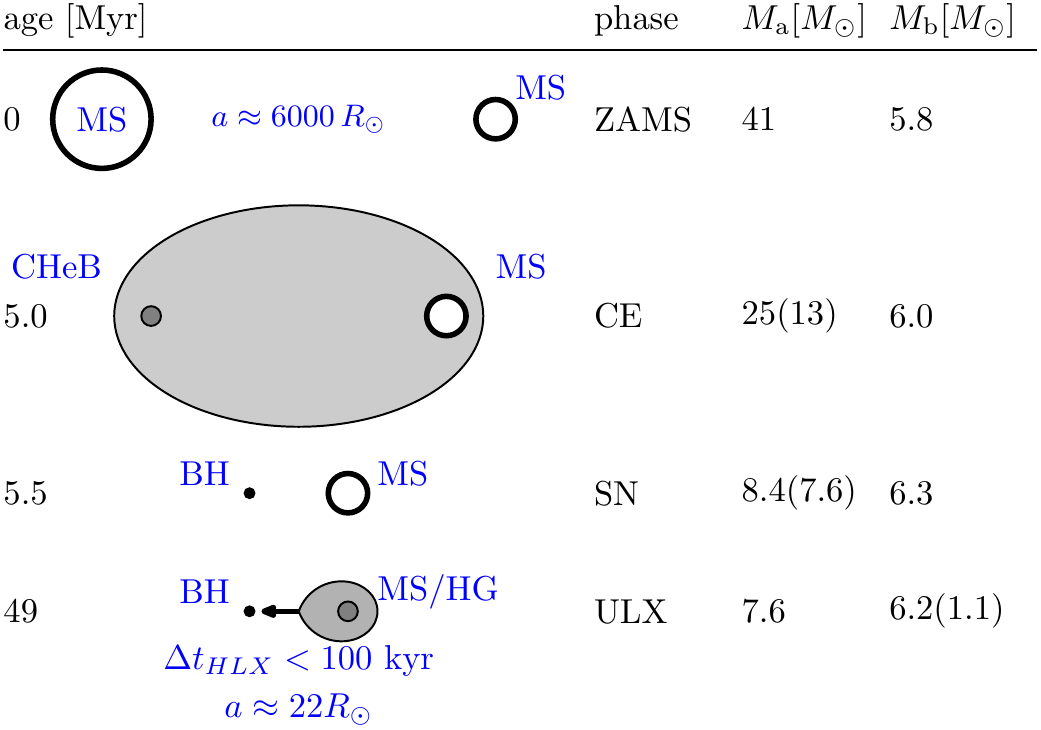}
    \rule{\columnwidth}{1pt}
    \includegraphics[width=0.9\columnwidth]{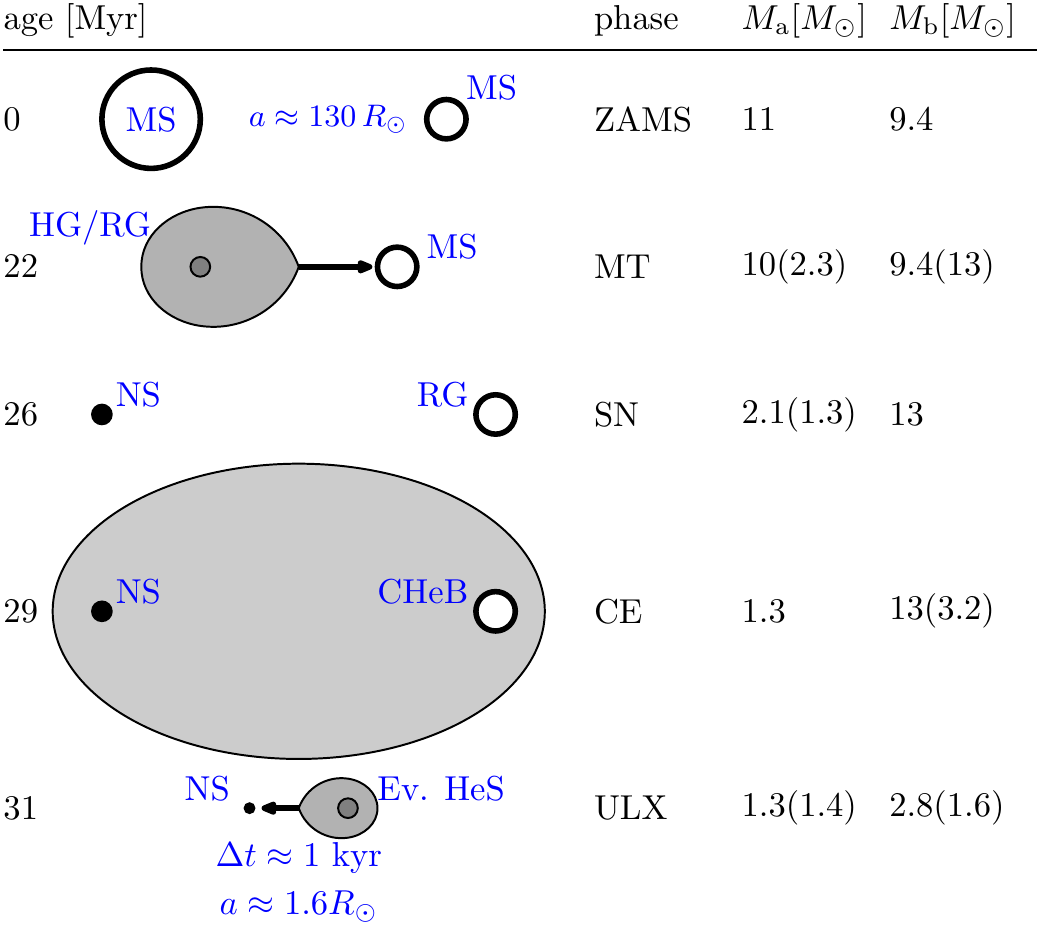}
    \caption{Similar as in Fig.~\ref{fig:rulx}, but for the evolution of the most luminous
        ($\lx>10^{41}\ergs$) ULXs.  Top/bottom plot is for a BH/NS accretor.}
    \label{fig:hlx}
\end{figure}

\paragraph{$\mathbf{15\dash60\myr}$; Route \rbhhg; Typical companion HG} A
typical system begins its evolution as $25\dash50\msun$ primary with a
$5\dash10\msun$ companion. After $\sim5\myr$ and the loss of $1\dash10\msun$ in
stellar wind the primary fills the RL as a CHeB star and commences the CE
phase.  It loses a large fraction of its mass ($\sim50\%$), but the separation
shrinks to $\sim15\rsun$. After additional $\sim0.5\myr$ a $\sim7\dash18\msun$
BH forms in a direct collapse. The secondary becomes a HG star
$\sim10\dash55\myr$ later and it expands rapidly. Shortly after, it fills its
RL and starts the MT. For a short time ($<0.1\myr$) the MT is high enough to
power a ULX with $\lx>10^{41}\ergs$ (see Fig.~\ref{fig:hlx}, top).

\paragraph{$\mathbf{17\dash75\myr}$; Route \rnshehg; Typical companion evHeS}
The stars on ZAMS are less massive and have smaller mass ratio than in the case
of \rbhhg. When the primary fills the RL as a RG, the MT is stable and proceeds
for a few tens of $\kyr$. The donor losses most of its hydrogen envelope and
after a few \myr s at the age of $\sim25\myr$ it forms a NS through a SN
explosion. The companion, being a RG, expands and fills its RL.  This time the
MT is unstable due to a high mass ratio and the CE occurs. Afterwards, the
secondary becomes a low-mass (a few Solar masses) helium star (HeS).  The
secondary continues its evolution for the next few \myr s, expands and fills
its RL again. This time, although being very strong, the MT is stable and for
$\sim1\kyr$ it powers an HLX (see Fig.~\ref{fig:hlx}, bottom).

\section{Discussion} \label{sec:discussion}

ULXs are observed in all kinds of galaxies \citep[e.g.][]{Mushotzky0601}.
However, ULXs appear to be more abundant in the star-forming galaxies than in
the elliptical galaxies \citep[e.g.][however see \citeauthor{Soria0704}
\citeyear{Soria0704}]{Irwin0402,Swartz0410,Feng1111,Wang1609}. In addition,
ULXs are commonly detected in star-burst regions
\citep[e.g.,][]{Fabbiano0106,Gao0310}. \citet{Feng1111} pointed out that this
connection is particularly valid for the most luminous ULXs. On the contrary,
\citet{Mushotzky0601} performed a direct examination of the ULXs' positions and
found that approximately $1/3$ of them had to form outside of the star-forming
regions.

These observations may be understood on the ground of our results.  On the one
hand, spatial correlation of ULXs and SF regions is expected because during the
SF episodes ULXs form mainly through routes \rbhms\ and \rnsms, which produce
numerous sources early after the start of SF.  Additionally, the most luminous
ULXs (HLXs) formed in our simulations evolve from heavier stars at an early age
($\tage\lesssim200\myr$).  On the other hand, the ULX phase in systems formed
outside the SF regions occurs later on after ZAMS (\rnshg, \rnsrg).  However,
we found that these late-time ULXs constitute only $~1\dash10\%$ of all ULXs
for constant SF.

\subsection{NS accretors in ULXs}

Our results predict a large fraction of NS accretors among the ULXs.  Current
observational status is consistent with our results.  The discovery of the
first NSULX \citep{Bachetti1410} proved that apparent luminosities orders of
magnitude larger than the Eddington limit are possible in nature.
\citet{Kluzniak1503} proposed that the M82 X-1 and the P13 in NGC 7793 may be
also PULXs. Following discoveries \citep{Furst1611,Israel1609} showed that NSs
may indeed be widespread among the ULXs. Finally, a known X-ray pulsar SMC X-3
was reported to reach an outburst luminosity of
\nergs{2.5}{39}\citep{Tsygankov1702}.  \citet{Pintore1701} argued that the
energy spectra of several ULXs can be explained with a 'pulsator-like'
continuum model used to describe Galactic XRBs containing NSs
\citep[e.g.][]{White95,Coburn0105}.

Theoretical considerations further support our findings.  \citet{King1605}
showed that a ULX with a weakly magnetized NS and a ULX with a BH may be
observationally hard to distinguish if the NS does not pulsate.
\citet{King1702} argued that the conditions for pulsations in PULXs are
relatively strict and a large population of NSULXs will mimic the BHULXs.

The significant number of NSULX that follows from most of our simulated models
may be understand on the statistical ground. While it is easier to obtain large
MT rates in the BHULXs, the NSULXs progenitors are more abundant in the ZAMS
distributions.  In general, NSs form from stars with ZAMS masses
$\sim8\dash22\msun$, whereas BH form from more massive stars with
$\sim22\dash150\msun$. For $\alpha_{IMF}=-2.7$ it gives the ratio of NS to BH
progenitors of $\sim4.8$, and for $\alpha_{IMF}=-2.3$ this ratio is $3.0$.
Moreover, in the NSULX route which dominates in the old stellar populations
(\rnsrg) the primary on ZAMS may have even lower mass of $\sim6\msun$, which
gives the ratio of NS to BH progenitors of $4.8$ for $\alpha_{IMF}=-2.3$. It
first becomes a WD before accreting mass and collapsing into a NS.
Additionally, donors in NSULXs are less massive than in BHULXs and, therefore,
also more abundant on ZAMS.  We note that low mass donors may occur also in
BHULXs, but it is far harder (and therefore less probable) for them to survive
the CE. Even if they succeeded, the post-CE separation is so small that the
RLOF occurs earlier than typically for NSULXs. All these results in a
significantly smaller number of BHULXs with low-mass ($\lesssim2\msun$) donors
than in NSULXs.

During short SF episodes, the ULX population is dominated by sources with BH
accretors. BHs may acquire stable MT from heavier companions than NSs. Heavier
stars expand more and faster, therefore, BHULXs age during ULX phase is of the
order of $\sim10\myr$, whereas for NSULX it is $\sim100\myr$ (route \rnsms). As
a result BHULXs appear in larger numbers in the first $\sim100\myr$.  The ULX
phase may appear in NSULX as early as $\sim10\myr$ after ZAMS. However, it
demands a very specific pre-CE conditions, which occur in a relatively narrow
ZAMS parameter space. For an average post-CE separation in early NSULXs
(\rnsms) the star needs $\sim100\myr$ to fill its RL.

The picture is different for the constant SF or long bursts. As the SF
continues, the \nbhulx\ saturates for $\sim200\myr$ after the SF start, because
at this time comparable number of sources begin and end their ULX phase.
However, a significant number of NSULXs have their ULX phases during the
post-MS expansion of their light donors. Therefore, the \nnsulx\ grows
continuously for  several {\gyr} and new NSULX formation routes appear
successively (see Sec.~\ref{sec:nsulx} for the sequence).  For $Z=\zsun$, the
\nnsulx\ becomes higher than \nbhulx\ after $\sim1\gyr$. However, for lower
metallicities, \nnsulx\ does not overcome \nbhulx\ within $10\gyr$.

When the SF is extinguished, \rbhms\ quickly disappear as their massive donors
end their lives fast ($\lesssim500\myr$). On the other hand, NSULXs with
low-mass donors may commence the ULX phase after several {\gyr} of evolution
when the secondary fills the RL during the post-MS expansion (\rnshg\ and
\rnsrg). Therefore, after the end of the SF, NSULXs quickly (in $\sim100\myr$)
become the dominant ULXs and after $\sim1\gyr$ BHULXs disappear completely.

\bigskip

The population synthesis of NSULXs was recently performed with the use of BSE
code by \citet{Fragos1503}. However, they omitted the pre-SN evolution and
explored a far smaller parameter space than that presented in this work.  Even
though they utilized the MESA code to calculate the MT rates, their maximum
values are similar to ours ($\mdot\lesssim10^{-2}\myr$). We obtained less
massive companions in NSULXs, but \citep{Fragos1503} focused on M82 X-2 source
with companion mass $\gtrsim5.2\msun$ \citep{Bachetti1410}  and assumed that
heavier stars are necessary to survive the CE phase.

\citet{Shao1504} also took the population synthesis approach to study NSULXs
and found a significant population of these sources.  They adopted a realistic
envelope's binding energy from \citet{Xu1010}, but considered only a solar
metallicity and constant beaming of $b=0.1$.

\subsection{Dependence of \nulx\ on Metallicity}\label{sec:disc_Z}

Observations show that ULXs are often associated with low metallicity
environments \citep[e.g.][]{Pakull0202,Soria0501,Luangtip1501,Mapelli1010}.
Many authors have demonstrated that in a low metallicity environment it is
easier to form a massive BH
\citep[e.g.,][]{Zampieri0912,Mapelli0905,Belczynski1005}, and it is more common
to obtain a RLOF onto a compact object \citep{Linden1012}. 

In this study we find that the metallicity has a strong impact on \nulx, but
only in SF regions (See Tab.~\ref{tab:num}). We show that the number of
\nbhulx\ increases significantly for \zsun/10 in comparison to \zsun. However,
the \nnsulx\ is virtually unaffected by metallicity. Interestingly, for models
with the lowest investigated metallicity ($\zsun/100$) we found fewer ULXs than
for $Z=\zsun/10$. Therefore, our results suggest that the relation $\nulx(Z)$
is not monotonic \citep[see][]{Prestwich1306}. We plan to study this outcome in
more details in a forthcoming publication (Wiktorowicz et al. in preparation).

\subsection{Counterparts}

Several searches were performed to directly observe the ULX companions.
However, only a handful have been detected among hundreds of known ULXs.
Additionally, ULXs, being extra-galactic objects, are observationally biased
towards more luminous and therefore more massive companions. In our results, NS
counterparts in ULX are predominantly low-mass ($\lesssim1.5\msun$) stars.  If
located in other galaxies, such objects would be hard to detect. This may
explain why it is hard to find counterparts to majority of the ULXs.

Some ULXs were observed near OB stars \citep[e.g.][]{Soria0501}. Such stars are
too massive to provide a stable MT to a NS accretor, so they may only be
accompanied by BH accretors, unless the MT occurs through a stellar wind (Not
considered in our study).  OB stars evolve quickly ($<100\myr$), which suggests
that they should be spatially associated with the SF regions, where the BHULXs
are more numerous than the NSULXs according to our results. Typical BHULX
companions in our simulations are MS stars of $5.6\dash11\msun$, which matches
the B spectral type.

\citet{Roberts0806} found 5 potential optical ULX companions with the use of
the Hubble Space Telescope. The optical search for companions in ULXs was
performed also by \citet{Gladstone1306} who used the same instrument and found
$13\pm5$ counterparts to ULXs located within a distance of $D\leq5\mpc$. Most
of them were MS stars or RGs. A near-infrared (NIR) search was performed by
\citet{Heida1408,Heida1606} who observed 62 ULXs in 37 galaxies located within
$10\mpc$. They found 17 candidate counterparts, 13 of which are red
supergiants. We note that red supergiants are present in our results in
small numbers ($<1\%$). The relatively large number of these stars in
observations may be an observational bias, as they are significantly brighter
than the most typical companions predicted by our study ($<2\msun$).
Nevertheless, red supergiants may be an interesting topic of a separate study.

The companion to NS ULX P13 in NGC7793 was detected and estimated to be a BI9a
star with mass $18\dash23\msun$ \citep{Motch1410}. Radius of such a star is
$4\dash700\rsun$, depending on its evolutionary stage.  Orbital period estimate
($P_{\rm orb}=64$ d) puts a Roche lobe radius of  a donor at $\sim120\rsun$
(under the assumption of circular orbit and $\mns=1.4\msun$).  The BI9a star
may obtain such a radius during CHeB phase.  According to a standard
prescription, in a high mass-ratio system the RLOF will be unstable and the CE
will occur.  However, \citet{Pavlovskii1702} found that the range of stable
mass ratios may be wider than previously thought. We will analyse this
possibility in a separate study (Wiktorowicz et al. in preparation). BHULXs
with high-mass donors may be predecessors of double BHs mergers
\citep{Belczynski1606}.

\section{Summary and Conclusions} 

We analyzed accretion onto compact stars in order to investigate the parameters
and characteristics of the ULX population. We demonstrates that on the grounds
of the current understanding of the accretion physics and stellar evolution we
{\it are} able to reproduce the vast majority of the observed population of
ULXs without the need for intermediate-mass black hole accretors.

We performed a population synthesis study with the use of \startrack\
population synthesis code to simulate \ttt{2}{7} isolated binaries with initial
parameters leading to the formation of XRBs (both LMXB and HMXB). Our
simulation grid involves various accretion models, beaming prescriptions, and
metallicities.

Our main conclusions are as follows:

\begin{itemize}
    \item {\it ULX with NS accretors dominate the post-burst ULX populations},
    and constant SF (duration $>1\gyr$) in high-$Z$ environments, what is a
    natural consequence of the current understanding of the binary evolution.
    NSULXs are present in significant numbers ($\gtrsim10\%$) also during the
    SF bursts and in lower-$Z$ ULX populations.
    
    \item {\it ULXs appear in a very specific sequence after the start of the SF}
        ($t$ depicts here the time since the beginning of the SF): 
        \begin{enumerate}
            \item \makebox[3cm][l]{$t\approx4\dash40\myr$} BH\dash MS
                ($5.6\dash11\msun$),
            \item \makebox[3cm][l]{$t\approx6\dash800\myr$}   NS\dash MS
                ($0.9\dash1.5\msun$),
            \item \makebox[3cm][l]{$t\approx430\dash1100\myr$} NS\dash HG
                ($0.6\dash1.0\msun$),
            \item \makebox[3cm][l]{$t\approx540\dash4400\myr$}  NS\dash RG
                ($\sim1.0\msun$);
        \end{enumerate}

    \item {\it We found that NSULXs may reach luminosities as high as those of
    BHULXs} ($\lxmax>\lhlx$).

    \item {\it The most luminous ULXs ($\lx\gtrsim\lhlx$) contain HG donors
    (BHULXs; $\mb\approx 1.2\dash3.7\msun$) or evolved helium stars. (NSULXs;
    $\mb\approx 1.7\dash2.6\msun$)}, which overfill their RL and transfer mass
    on thermal-timescale. They form typically within $15\dash75\myr$ since
    ZAMS.

    \item {\it The number of ULXs is anti-correlated with metallicity}.
    However, for very low-$Z$ this relation changes sign.

\end{itemize}

\acknowledgements
We would like to thank volunteers whose participation in the Universe@Home
project\footnote{http://universeathome.pl} made it possible to acquire the
results in such a short time and the anonymous referee who helped to
significantly improve the paper. GW and KB would like to thank W. Kluzniak for
interesting discussions.  This study was partially supported by the Polish
National Science Center NCN grants: N203 404939, OPUS 2015/19/B/ST9/01099, and
SONATA BIS 2 DEC-2012/07/E/ST9/01360.  MS was supported by NASA contract
NAS8-03060 (Chandra X-ray Center). MS acknowledges also the Polish NCN grant
OPUS 2014/13/B/ST9/00570.  JPL was supported in part by a grant from the French
National Space Center CNES.  KB acknowledges support from the NCN grants: OPUS
2015/19/B/ST9/03188 and MAESTRO 2015/18/A/ST9/00746.  GW was supported by the
NCN grant OPUS 2015/19/B/ST9/03188 and HARMONIA 2014/14/M/ST9/0707.  This
research was supported in part by the National Science Foundation under Grant
No. NSF PHY-1125915.

\bibliographystyle{apj}
\bibliography{ms}

\appendix

\section{Details of the simulations}

If not stated differently, throughout the article a 'luminosity' means an X-ray
luminosity in the $0.2\dash10$ keV band. We do not take into account NS
magnetic fields and include only non-rotating BHs. All of the frequently used abbreviations are
summarized in Tab.~\ref{tab:abbreviations}.

We utilized the EL for luminosity (\ledd) and mass accretion (\medd) as
\begin{equation}\label{eq:ledd}
    \ledd=\eta\medd
    c^2=\nergs{2.51}{38}\frac{1}{1+X}\left(\frac{\macc}{\msun}\right),
\end{equation}

\noindent where $\eta=1/12$ for BH accretors and $\sim0.2$ for NSs, $X$ is a
fraction of Hydrogen in the envelope of the donor. $X=0.7$ for hydrogen reach
donors and $X=0$ for HeS and WD.

Most of the results are provided as present time distributions. This means the
distributions as they will be visible by an observer in present time ($10\gyr$
after the SF beginning for constant SF, or ($100\myr$, $1\gyr$, $5\gyr$,
$10\gyr$ after the SF beginning for burst SF). We omit issues related to
obscuration or light propagation time.

Below we list the additional models tested in our grid of simulations. The
standard model will be referred to as AD1 BK.

\begin{deluxetable}{ll}
    \tablecaption{Frequently used abbreviations}
    \startdata
    \hline\\
    ULX & UltraLuminous X-ray source\\
    BHULX & ULX with a BH accretor \\
    NSULX & ULX with a NS accretor \\
    PULX & Pulsing ULX\\
    MS  & Main Sequence\\
    HG & H-rich Hertzsprung Gap\\
    RG & Red Giant\\
    CHeB & Core Helium Burning\\
    HeS & Helium Star\\
    evHeS & evolved Helium Star\\
    SN & SuperNova\\
    AIC & Accretion Induced Collapse\\
    RL & Roche Lobe\\
    RLOF & Roche Lobe OverFlow\\
    MT & Mass Transfer\\
    GR & Gravitational Radiation\\
    SF & Star Formation\\
    WD & White Dwarf\\
    ONeWD & Oxygene-Neon WD\\

    \enddata
    \label{tab:abbreviations}

\end{deluxetable}

\subsection{The ''upper limit'' accretion model (AD0)}\label{sec:acc_model_ad0}

In addition to accretion model described in Sec.~\ref{sec:acc_model}, to which
we will refer to as {\bf AD1}, we investigated a possibility that all mass may
be transferred from the donor and efficiently accreated by the accretor, i.e. 

\begin{equation}
    \macc=\mrlof.
\end{equation}

This corresponds to the highest accretion rate that could potentially occur in
a given system, but is most likely not realistic for high \mrlof. Therefore,
this model should be considered as a rough upper limit for accretion.

The X-ray luminosity is calculated as

\begin{equation}\label{eq:lx} 
    \lx = \frac{\epsilon G \mbhns \; \mrlof}{\racc} = \eta\mrlof c^2, 
\end{equation} 

\noindent where $\epsilon$ is a conversion efficiency of gravitational energy
into radiation equal $1.0$ for a NS (surface accretion) and $0.5$ for a BH
(disk accretion), $\mbhns$ is the mass of the accretor, $\racc$ is the radius
of a NS (assumed to be $10$ km) or a BH (3 Schwarzschild radii, non-rotating
BH), and $\eta$ is is the radiative efficiency of a standard thin disk equal
$\sim0.2$ for a NS and $1/12$ for a BH \citep[e.g.,][]{Shakura73}.

\subsection{Models of the beaming}\label{sec:beaming} 

In addition to the beaming model described in Sec.~\ref{sec:beaming_model} to
which we will refer to as {\bf BK}, we investigated also the following models:

\subsubsection{No beaming ({\bf BN})}

For this model we assumed isotropic emission, therefore,

\begin{equation}
    b=1
\end{equation}

\noindent for all systems and all mass transfer rates.

\subsubsection{Constant beaming, no saturation ({\bf B01})}\label{sec:beaming_const}

We start with the simplest model of constant beaming, and we apply it to all
sources. We consider only one case of $b=0.1\, (\theta\approx52^\circ)$  to
compare it with more realistic prescriptions described below. Such a constant
beaming will lower the total luminosity which is required for a source to be
observed as a ULX to $b\times\lulx$. This will generally increase the number of
predicted ULX sources. On the other hand, \pobsb\ will be lower for all
systems, which will decrease this number. These two processes may make the
predicted number of ULXs lower or higher in comparison to models without
beaming.

\subsubsection{S\k{a}dowski's model with saturation ({\bf BS})}\label{sec:beaming_bs}

\begin{figure}[h]
    \centering
    \includegraphics[scale=0.5]{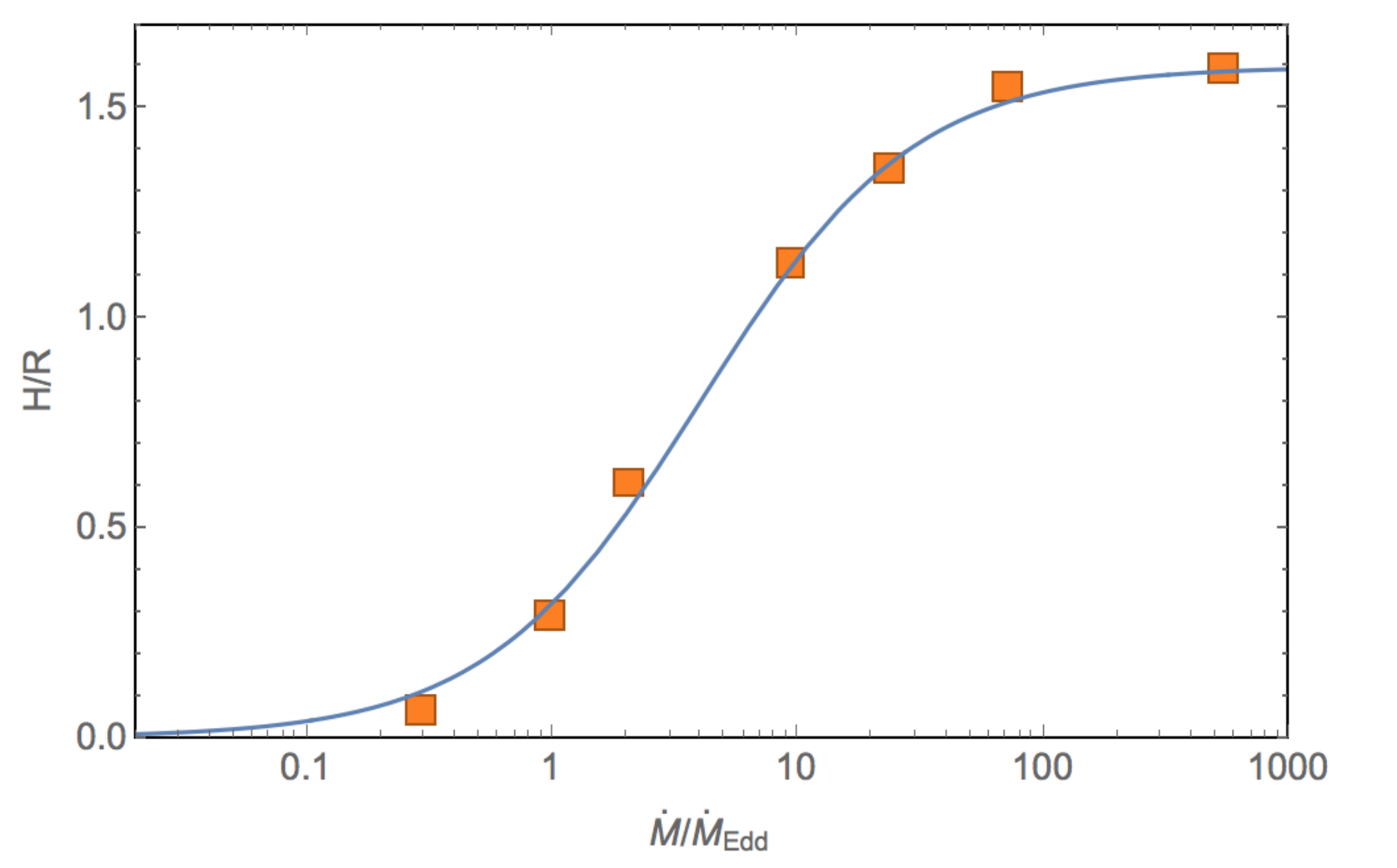}
    \caption{The ratio of photosphere height to disk radius ($H/R$) in relation
    to the mass transfer rate (in Eddington units). Orange squares represent
    the results of GRRMHD simulations with the use of {\tt KORAL} code
    (\citeauthor{Sadowski1403}, private communication). The blue line shows the
    fit to the data (Eq.~\ref{eq:bs}).} \label{fig:bs}
\end{figure}

This beaming prescription is based mainly on the results of theoretical and
numerical analysis presented in \citet{Lasota1603} and based on the results of
{\tt KORAL} \citep{Sadowski1403} GRRMHD simulations. They showed that
super-Eddington disks never become geometrically thick, because the thickness
of slim disk does not depend on the MT rate. Even for very high \mt\ the ratio
of photosphere height ($H$) to radius ($R$) is $H/R\lesssim1.6$.  We fitted a
phenomenological model to approximate the relation between $H/R$ and \mt\ found
by \citet{Lasota1603} (Fig.~\ref{fig:bs}), and we obtained

\begin{equation}\label{eq:bs}
    \frac{H}{R}=\frac{1.6}{1+\frac{4}{\mt}}
\end{equation}

\noindent for non-rotating BH accretion with non-saturated magnetic field. This
relation corresponds to $R=30\rg$, but $H/R$ should not be significantly larger
for other radii. The equation shows a moderate photosphere height also for
sub-Eddington MT rates. We used the same prescription for NS accretors.
 
The $H/R$ is related with the opening angle $\theta$ as

\begin{equation}\label{eq:bs_hr}
    \left(\frac{H}{R}\right)^{-1}=\tan\frac{\theta}{2},
\end{equation}

\noindent which allows to derive the beaming factor as

\begin{equation}
    b=1-\cos\frac{\theta}{2}
\end{equation}

\noindent and the apparent isotropic luminosity is calculated as in Eq.~\ref{eq:lxiso}.

The beaming becomes nearly constant for $\mt\gtrsim100$ what give the minimum
opening angle $\theta_{\rm min}\approx64^\circ$ ($b\approx0.15$). The
prescription works for the entire range of \mt\ and shows a small collimation
also for nearly-Eddington MT ($\mt<1$). However, below $\mt\approx0.12$ beaming
factor is always greater than $b>0.95\,(\theta>170^\circ)$.

\section{Additional Characteristics of the Simulated Populations
of ULXs}

\begin{figure}[h]
    \centering
    \includegraphics[width=0.99\textwidth]{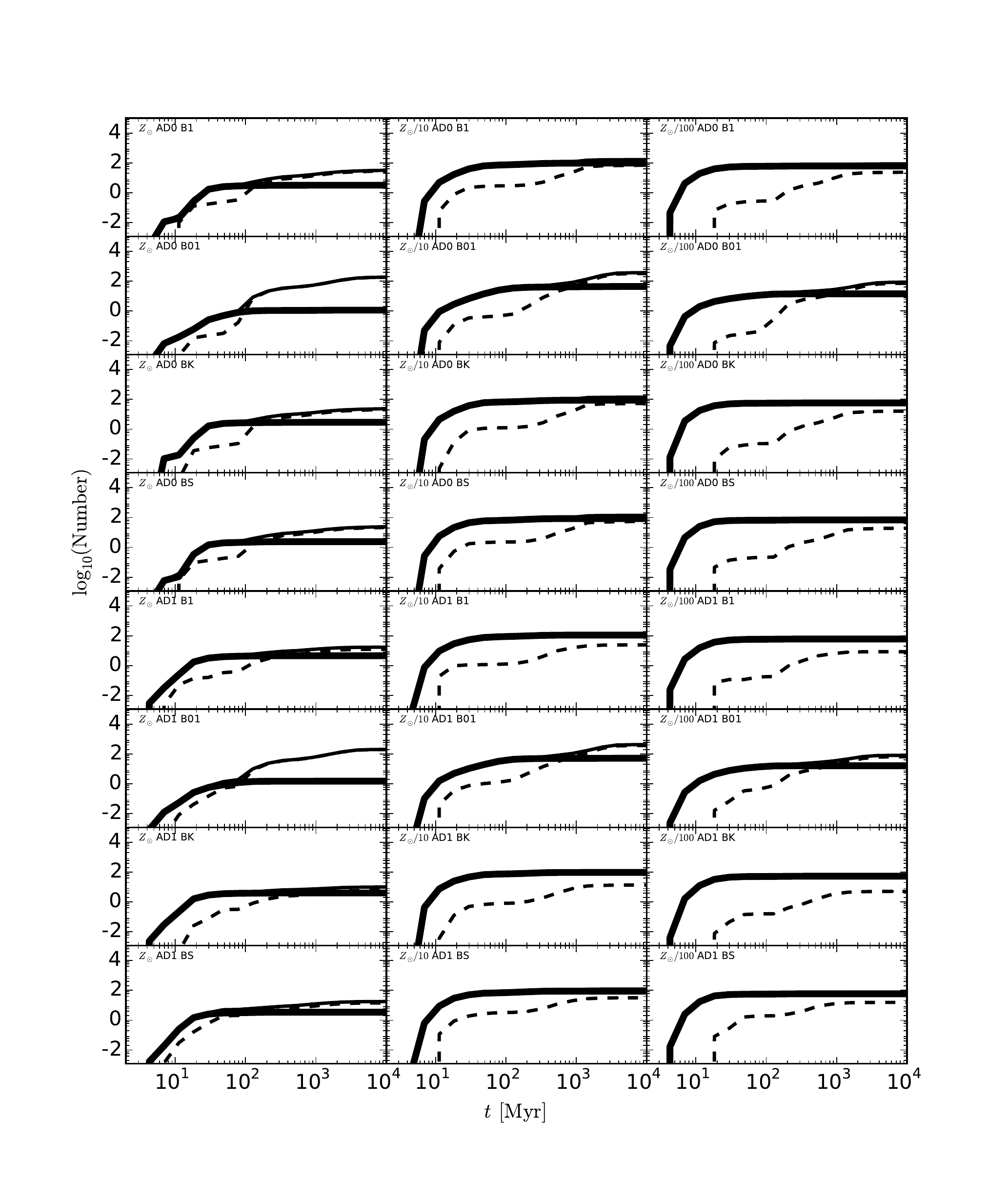}
    \caption{The relation between the number of ULXs and the age of the
    population for constant SFR of $6.0\msy$ (total stellar mass equal
    Milky-Way Equivalent Galaxy). The thick line corresponds to BHULXs, dashed
    to NSULXs, and solid to all ULXs. $\zsun$ depicts the solar metallicity,
    AD0/1 are the accretion models (Secs.  \ref{sec:acc_model_ad0} and
    \ref{sec:acc_model}) and BN,B01,BK,BS stand for beaming models
    (Secs.~\ref{sec:beaming} and \ref{sec:beaming_model}).}
        \label{fig:num_dc_grid}
\end{figure}

\begin{figure}[h]
    \centering
    \includegraphics[width=0.99\textwidth]{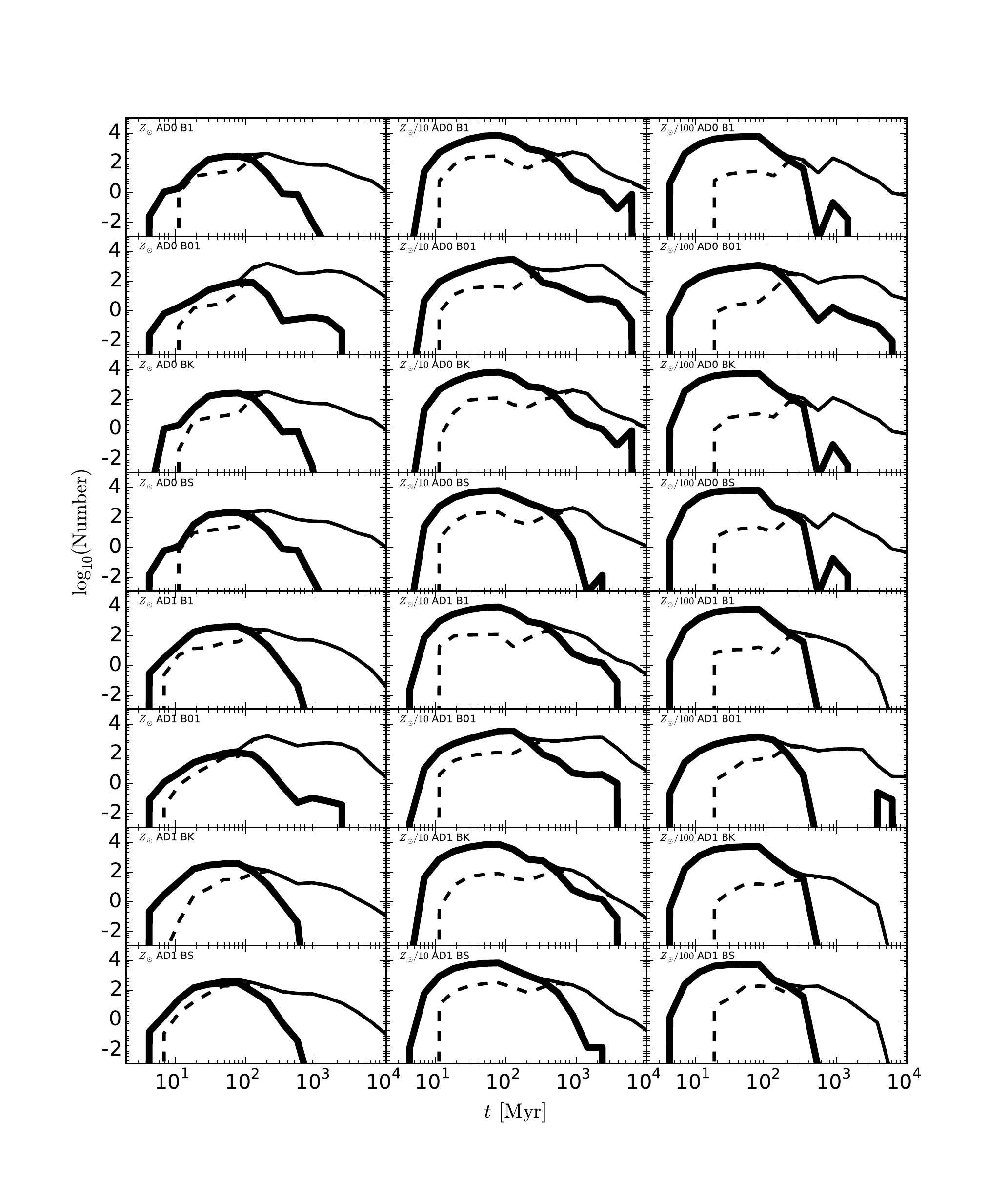}
    \caption{The same as in Fig.~\ref{fig:num_dc_grid}, but for burst
    star-formation with a duration of $100\myr$ and SFR of $600\msy$.}
    \label{fig:num_d100_grid}
\end{figure}

Tab.~\ref{tab:num_sfb_lxmin} contains the dependence of \nulx\ on \lxmin. The distributions of the general population’s parameters are in Tabs.~\ref{tab:results_sfc}, \ref{tab:results_sf100}, \ref{tab:results_sf1000}, \ref{tab:results_sf5000}, and \ref{tab:results_sf10000}. The evolution of the \nulx\ through population age are in Figs.~~\ref{fig:num_dc_grid} and \ref{fig:num_d100_grid}. Tabs.~\ref{tab:comps_sfc}, \ref{tab:comps_sf100}, \ref{tab:comps_sf1000}, \ref{tab:comps_sf5000}, and \ref{tab:comps_sf10000} contain the distributions of companions.

\newpage
\begin{figure}[h]
    \centering
    \includegraphics[width=0.49\textwidth]{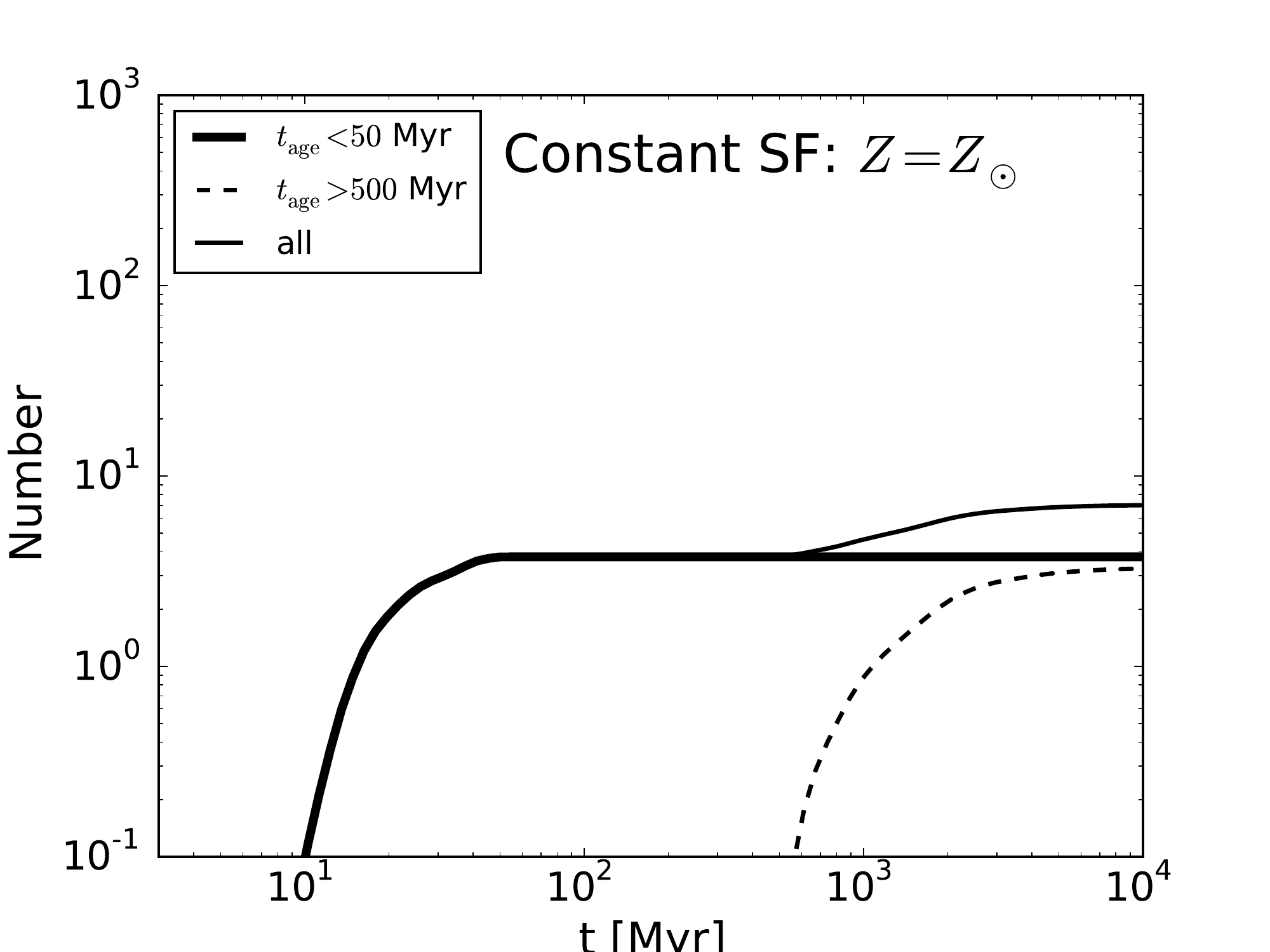}
    \includegraphics[width=0.49\textwidth]{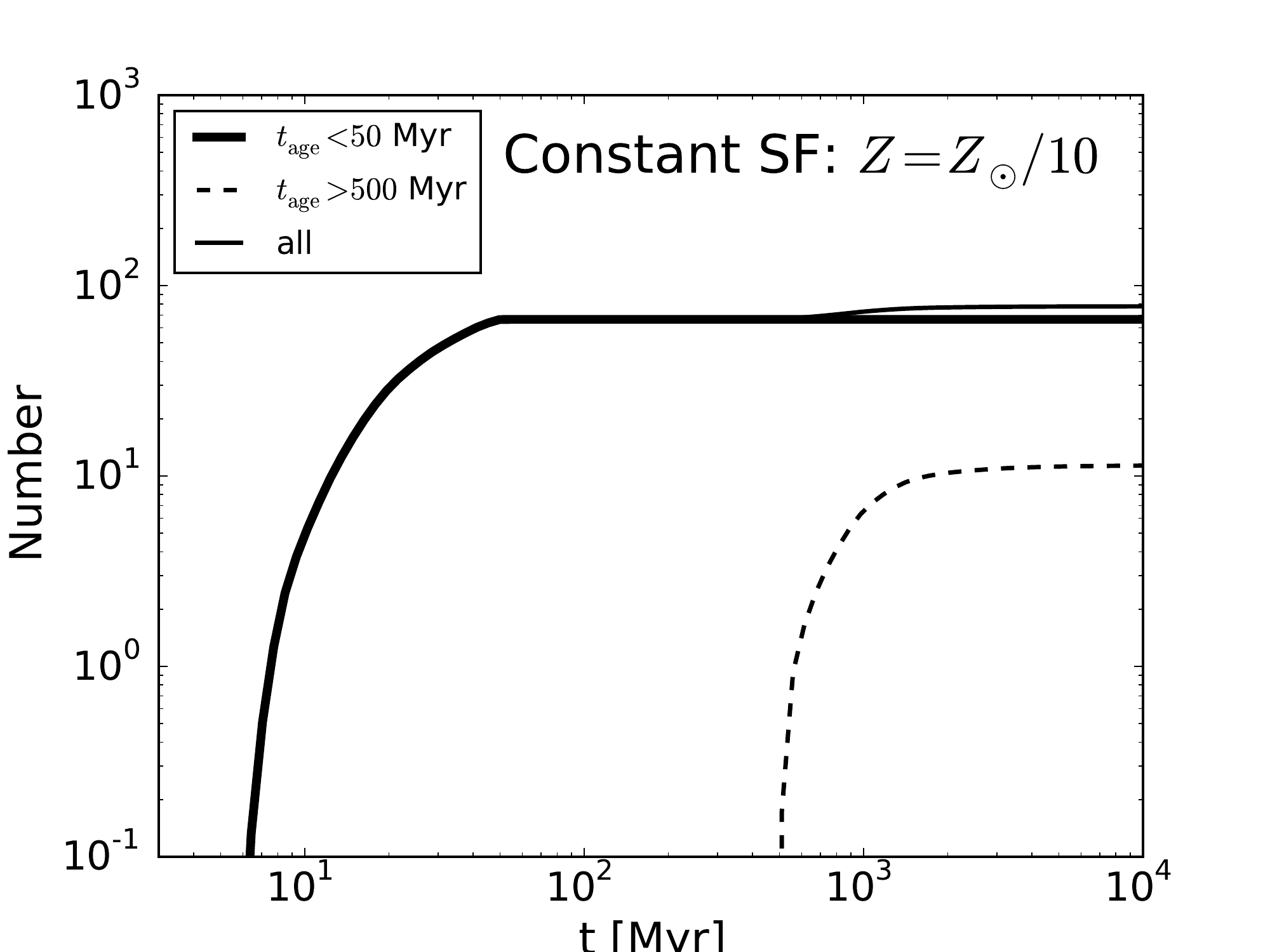}
    \caption{The evolution of the relative abundance of young ($t_{\rm
    age}<50\myr$; mainly routes \rbhms\ and \rnsms) and old ($t_{\rm
    age}>500\myr$; mainly routes \rnshg\ and \rnsrg) ULXs through time for
    constant star-formation rate and two metallicities: $Z=\zsun$ (left) and
    $Z=\zsun/10$ (right).}
    \label{fig:num_yo}
\end{figure}

\begin{deluxetable*}{llccccc}
    \tablewidth{\textwidth}
    \tablecaption{\nulx\ for burst and constant SFR)}
    \tablehead{& & \multicolumn{5}{c}{Time since the start of star-formation}\\ & & \multicolumn{4}{c}{burst SFR\tablenotemark{a}} & constant SFR\\
            Metallicity & & 100\myr & 1\gyr & 5\gyr & 10\gyr & 10\gyr }
    \startdata
          & \nulx   & \ttt{4.0}{2}      & \ttt{1.8}{1}  & \ttt{9.5}{-1} & \ttt{1.1}{-1} & \ttt{9.3}{0}\\
    \zsun & \nbhulx & \ttt{3.7}{2}      & \ttt{2.8}{-7} & --            & --            & \ttt{3.8}{0}\\
          & \nnsulx &  \ttt{3.5}{1}     & \ttt{1.8}{1}  & \ttt{9.5}{-1} & \ttt{1.1}{-1} & \ttt{5.5}{0}  \\ 

          & \nulx   & \ttt{7.3}{3}      & \ttt{1.1}{2}  & \ttt{7.0}{-1} & \ttt{7.5}{-2} & \ttt{1.0}{2} \\ 
    \zsun/10 & \nbhulx & \ttt{7.2}{3}   & \ttt{4.0}{0}  & --            & --            & \ttt{9.2}{1}  \\ 
          & \nnsulx & \ttt{8.1}{1}      & \ttt{1.1}{2}  & \ttt{7.0}{-1} & \ttt{7.5}{-2} & \ttt{1.3}{1}\\
           
          & \nulx   &\ttt{5.0}{3}       & \ttt{2.9}{1}  & \ttt{1.0}{-1} & \ttt{5.0}{-7} & \ttt{5.6}{1} \\ 
    \zsun/100 & \nbhulx & \ttt{4.9}{3}  & --            & --            & --            & \ttt{5.1}{1} \\ 
          & \nnsulx & \ttt{1.6}{1}      & \ttt{2.9}{1}  & \ttt{1.0}{-1} & \ttt{5.0}{-7} & \ttt{5.0}{0}\\ 
    \enddata
 \tablecomments{Number of ULXs (\nulx) in the reference model with a division
 on BHULXs (\nbhulx) and NSULXs (\nnsulx) for different population ages. BHULX
 are present also late after the burst ($\sim1\gyr$), but \nbhulx\ is
 negligible in that population's age in comparison to \nnsulx\ or the \nbhulx\
 during the burst.} \tablenotetext{a}{Burst duration is $100\myr$}
\label{tab:num}
\end{deluxetable*}

\begin{deluxetable*}{llcccc}
    \tablewidth{\textwidth}
    \tablecaption{\nulx(\lxmin)  $100\myr$ after SF start}
    \tablehead{& & \multicolumn{4}{c}{\lxmin}\\
    Metallicity &  & $10^{39}\ergs$ & \nergs{3}{39} & $10^{40}\ergs$ & $10^{41}\ergs$ }
    \startdata
                & \nulx   &  \ttt{4.0}{2}  &  \ttt{6.5}{1}  &  \ttt{1.4}{1}  &  \ttt{5.0}{-1} \\
    \zsun       & \nbhulx &  \ttt{3.7}{2}  &  \ttt{4.0}{1}  &  \ttt{4.7}{0}  &  \ttt{3.8}{-1} \\
                & \nnsulx &  \ttt{3.5}{1}  &  \ttt{2.5}{1}  &  \ttt{9.4}{0}  &  \ttt{1.3}{-1} \\
                \\
                & \nulx   &  \ttt{7.3}{3}  &  \ttt{1.9}{3}  &  \ttt{2.0}{2}  &  \ttt{1.2}{1} \\
    \zsun/10    & \nbhulx &  \ttt{7.2}{3}  &  \ttt{1.9}{3}  &  \ttt{1.8}{2}  &  \ttt{1.1}{1} \\
                & \nnsulx &  \ttt{8.1}{1}  &  \ttt{4.0}{1}  &  \ttt{1.2}{1}  &  \ttt{4.3}{-1}\\
                \\
                & \nulx   &  \ttt{5.0}{3}  &  \ttt{2.4}{3}  &  \ttt{3.9}{2}  &  \ttt{1.5}{1} \\
    \zsun/100   & \nbhulx &  \ttt{4.9}{3}  &  \ttt{2.4}{3}  &  \ttt{3.8}{2}  &  \ttt{1.5}{1} \\
                & \nnsulx &  \ttt{1.6}{1}  &  \ttt{1.2}{1}  &  \ttt{9.0}{0}  &  \ttt{7.5}{-2} \\
    \enddata
 \tablecomments{Number of ULXs depending on the minimal luminosity (\lxmin) for
 different metallicities with a division on BHULXs and NSULXs for AD1 BK
 model.}
\label{tab:num_sfb_lxmin}
\end{deluxetable*}

\begin{deluxetable*}{lccccc}[b]
    \tablewidth{\textwidth}
    \tablecaption{Number of ULXs and maximal luminosity (constant SFR\tablenotemark{a})}
    \tablehead{Model\tablenotemark{b} & \multicolumn{3}{c}{Number per
        MWEG\tablenotemark{c}} & \multicolumn{2}{c}{$\lxmax
        [\ergs]$\tablenotemark{d}} \\
& \nulx & \nbhulx & \nnsulx & BHULX & NSULX }
    \startdata
 & \multicolumn{5}{c}{$\zsun$}\\
AD0 BN     & \ttt{3.1}{1} & \ttt{3.1}{0} \hfill ($10\%$) & \ttt{2.8}{1} \hfill ($89\%$) & \ttt{8.4}{44} & \ttt{3.3}{44} \\
AD0 B01    & \ttt{1.8}{2} & \ttt{1.1}{0} \hfill ($0\%$) & \ttt{1.8}{2} \hfill ($99\%$) & \ttt{8.4}{45} & \ttt{3.3}{45} \\
AD0 BK     & \ttt{2.2}{1} & \ttt{2.7}{0} \hfill ($12\%$) & \ttt{1.9}{1} \hfill ($87\%$) & \ttt{2.6}{47} & \ttt{1.0}{47} \\
AD0 BS     & \ttt{2.0}{1} & \ttt{2.3}{0} \hfill ($11\%$) & \ttt{1.8}{1} \hfill ($88\%$) & \ttt{5.5}{45} & \ttt{2.2}{45} \\
AD1 BN     & \ttt{1.5}{1} & \ttt{4.4}{0} \hfill ($29\%$) & \ttt{1.1}{1} \hfill ($70\%$) & \ttt{3.0}{40} & \ttt{5.1}{39} \\
AD1 B01    & \ttt{2.0}{2} & \ttt{1.5}{0} \hfill ($0\%$) & \ttt{2.0}{2} \hfill ($99\%$) & \ttt{3.0}{41} & \ttt{5.1}{40} \\
AD1 BK     & \ttt{9.3}{0} & \ttt{3.8}{0} \hfill ($41\%$) & \ttt{5.5}{0} \hfill ($58\%$) & \ttt{9.2}{42} & \ttt{1.6}{42} \\
AD1 BS     & \ttt{2.0}{1} & \ttt{2.9}{0} \hfill ($14\%$) & \ttt{1.7}{1} \hfill ($85\%$) & \ttt{2.0}{41} & \ttt{3.4}{40} \\
 & \multicolumn{5}{c}{$\zsun/10$}\\
AD0 BN     & \ttt{1.6}{2} & \ttt{9.4}{1} \hfill ($58\%$) & \ttt{6.7}{1} \hfill ($41\%$) & \ttt{6.8}{45} & \ttt{2.3}{44} \\
AD0 B01    & \ttt{3.7}{2} & \ttt{4.3}{1} \hfill ($11\%$) & \ttt{3.3}{2} \hfill ($88\%$) & \ttt{6.8}{46} & \ttt{2.3}{45} \\
AD0 BK     & \ttt{1.3}{2} & \ttt{8.4}{1} \hfill ($63\%$) & \ttt{5.0}{1} \hfill ($36\%$) & \ttt{2.1}{48} & \ttt{7.1}{46} \\
AD0 BS     & \ttt{1.1}{2} & \ttt{6.4}{1} \hfill ($60\%$) & \ttt{4.1}{1} \hfill ($39\%$) & \ttt{4.5}{46} & \ttt{1.5}{45} \\
AD1 BN     & \ttt{1.3}{2} & \ttt{1.1}{2} \hfill ($83\%$) & \ttt{2.1}{1} \hfill ($16\%$) & \ttt{8.6}{40} & \ttt{4.5}{39} \\
AD1 B01    & \ttt{4.2}{2} & \ttt{5.0}{1} \hfill ($11\%$) & \ttt{3.7}{2} \hfill ($88\%$) & \ttt{8.6}{41} & \ttt{4.5}{40} \\
AD1 BK     & \ttt{1.0}{2} & \ttt{9.2}{1} \hfill ($87\%$) & \ttt{1.3}{1} \hfill ($12\%$) & \ttt{2.7}{43} & \ttt{1.4}{42} \\
AD1 BS     & \ttt{1.1}{2} & \ttt{6.8}{1} \hfill ($61\%$) & \ttt{4.2}{1} \hfill ($38\%$) & \ttt{5.7}{41} & \ttt{3.0}{40} \\
 & \multicolumn{5}{c}{$\zsun/100$}\\
AD0 BN     & \ttt{8.5}{1} & \ttt{6.1}{1} \hfill ($72\%$) & \ttt{2.3}{1} \hfill ($27\%$) & \ttt{8.9}{45} & \ttt{2.5}{44} \\
AD0 B01    & \ttt{8.6}{1} & \ttt{1.4}{1} \hfill ($16\%$) & \ttt{7.2}{1} \hfill ($83\%$) & \ttt{8.9}{46} & \ttt{2.5}{45} \\
AD0 BK     & \ttt{7.2}{1} & \ttt{5.6}{1} \hfill ($78\%$) & \ttt{1.6}{1} \hfill ($21\%$) & \ttt{2.7}{48} & \ttt{7.7}{46} \\
AD0 BS     & \ttt{4.9}{1} & \ttt{3.8}{1} \hfill ($78\%$) & \ttt{1.1}{1} \hfill ($21\%$) & \ttt{5.9}{46} & \ttt{1.6}{45} \\
AD1 BN     & \ttt{6.6}{1} & \ttt{5.9}{1} \hfill ($89\%$) & \ttt{7.2}{0} \hfill ($10\%$) & \ttt{1.3}{41} & \ttt{5.5}{39} \\
AD1 B01    & \ttt{8.0}{1} & \ttt{1.6}{1} \hfill ($19\%$) & \ttt{6.4}{1} \hfill ($80\%$) & \ttt{1.3}{42} & \ttt{5.5}{40} \\
AD1 BK     & \ttt{5.6}{1} & \ttt{5.1}{1} \hfill ($91\%$) & \ttt{5.0}{0} \hfill ($8\%$) & \ttt{4.0}{43} & \ttt{1.7}{42} \\
AD1 BS     & \ttt{5.0}{1} & \ttt{3.7}{1} \hfill ($73\%$) & \ttt{1.3}{1} \hfill ($26\%$) & \ttt{8.6}{41} & \ttt{3.6}{40} \\

 \enddata
\label{tab:results_sfc}
\tablenotetext{a}{At the age of 10 \gyr}
\tablenotetext{b}{AD0/1 - ``upper limit''/logarythmic accretion model (Secs.
    \ref{sec:acc_model_ad0} and \ref{sec:acc_model}); BN/B01/BK/BS - beaming
    models (Secs. \ref{sec:beaming} and \ref{sec:beaming_model}).}
\tablenotetext{c}{Mikly-Way Equivalent Galaxy}
\tablenotetext{d}{The maximal obtained luminosity for the particular accretor. In all cases the beaming which corresponds to this luminosities is saturated, so the beaming factor $b=1$, $0.1$, $\sim\ttt{3.2}{-3}$, and $\sim0.15$ for BN, B01, BK, and BS, respectively. Corresponding opening angles are $180^\circ$, $\sim52^\circ$, $\sim9^\circ$, and $\sim64^\circ$, respectively. }
\end{deluxetable*}

\begin{deluxetable*}{lccccc}
    \tablewidth{\textwidth}
    \tablecaption{General parameters of the simulational results (burst SFR $100\myr$ ago)}
    \tablehead{Model\tablenotemark{a} & \multicolumn{3}{c}{Number per
        MWEG\tablenotemark{a}} & \multicolumn{2}{c}{$\lxmax
        [\ergs]$\tablenotemark{b}} \\
& \nulx & \nbhulx & \nnsulx & BHULX & NSULX }
    \startdata
 & \multicolumn{5}{c}{$\zsun$}\\
AD0 BN     & \ttt{3.4}{2} & \ttt{2.8}{2} \hfill ($83\%$) & \ttt{5.8}{1} \hfill ($16\%$) & \ttt{8.4}{44} & \ttt{3.3}{44} \\
AD0 B01    & \ttt{4.7}{2} & \ttt{9.3}{1} \hfill ($19\%$) & \ttt{3.8}{2} \hfill ($80\%$) & \ttt{8.4}{45} & \ttt{3.3}{45} \\
AD0 BK     & \ttt{2.8}{2} & \ttt{2.6}{2} \hfill ($90\%$) & \ttt{2.8}{1} \hfill ($9\%$) & \ttt{2.6}{47} & \ttt{1.0}{47} \\
AD0 BS     & \ttt{2.8}{2} & \ttt{2.1}{2} \hfill ($75\%$) & \ttt{6.8}{1} \hfill ($24\%$) & \ttt{5.5}{45} & \ttt{2.2}{45} \\
AD1 B01    & \ttt{5.7}{2} & \ttt{1.3}{2} \hfill ($22\%$) & \ttt{4.4}{2} \hfill ($77\%$) & \ttt{3.0}{41} & \ttt{5.1}{40} \\
AD1 BK     & \ttt{4.0}{2} & \ttt{3.7}{2} \hfill ($91\%$) & \ttt{3.5}{1} \hfill ($8\%$) & \ttt{9.2}{42} & \ttt{1.6}{42} \\
AD1 BS     & \ttt{3.8}{2} & \ttt{2.8}{2} \hfill ($75\%$) & \ttt{9.4}{1} \hfill ($24\%$) & \ttt{2.0}{41} & \ttt{3.4}{40} \\
 & \multicolumn{5}{c}{$\zsun/10$}\\
AD0 BN     & \ttt{7.5}{3} & \ttt{7.2}{3} \hfill ($96\%$) & \ttt{2.9}{2} \hfill ($3\%$) & \ttt{6.8}{45} & \ttt{1.2}{44} \\
AD0 B01    & \ttt{3.0}{3} & \ttt{3.0}{3} \hfill ($98\%$) & \ttt{4.9}{1} \hfill ($1\%$) & \ttt{6.8}{46} & \ttt{1.2}{45} \\
AD0 BK     & \ttt{6.6}{3} & \ttt{6.4}{3} \hfill ($98\%$) & \ttt{1.2}{2} \hfill ($1\%$) & \ttt{2.1}{48} & \ttt{3.7}{46} \\
AD0 BS     & \ttt{5.2}{3} & \ttt{5.1}{3} \hfill ($98\%$) & \ttt{6.7}{1} \hfill ($1\%$) & \ttt{4.5}{46} & \ttt{7.9}{44} \\
AD1 BN     & \ttt{8.6}{3} & \ttt{8.5}{3} \hfill ($98\%$) & \ttt{1.0}{2} \hfill ($1\%$) & \ttt{8.6}{40} & \ttt{4.5}{39} \\
AD1 B01    & \ttt{3.9}{3} & \ttt{3.8}{3} \hfill ($96\%$) & \ttt{1.4}{2} \hfill ($3\%$) & \ttt{8.6}{41} & \ttt{4.5}{40} \\
AD1 BK     & \ttt{7.3}{3} & \ttt{7.2}{3} \hfill ($98\%$) & \ttt{8.1}{1} \hfill ($1\%$) & \ttt{2.7}{43} & \ttt{1.4}{42} \\
AD1 BS     & \ttt{5.7}{3} & \ttt{5.6}{3} \hfill ($97\%$) & \ttt{1.5}{2} \hfill ($2\%$) & \ttt{5.7}{41} & \ttt{3.0}{40} \\
 & \multicolumn{5}{c}{$\zsun/100$}\\
AD0 BN     & \ttt{5.9}{3} & \ttt{5.9}{3} \hfill ($99\%$) & \ttt{2.8}{1} \hfill ($0\%$) & \ttt{8.9}{45} & \ttt{2.1}{43} \\
AD0 B01    & \ttt{1.3}{3} & \ttt{1.3}{3} \hfill ($99\%$) & \ttt{4.4}{0} \hfill ($0\%$) & \ttt{8.9}{46} & \ttt{2.1}{44} \\
AD0 BK     & \ttt{5.4}{3} & \ttt{5.4}{3} \hfill ($99\%$) & \ttt{1.1}{1} \hfill ($0\%$) & \ttt{2.7}{48} & \ttt{6.5}{45} \\
AD0 BS     & \ttt{3.7}{3} & \ttt{3.7}{3} \hfill ($99\%$) & \ttt{6.9}{0} \hfill ($0\%$) & \ttt{5.9}{46} & \ttt{1.4}{44} \\
AD1 BN     & \ttt{5.7}{3} & \ttt{5.7}{3} \hfill ($99\%$) & \ttt{1.4}{1} \hfill ($0\%$) & \ttt{1.3}{41} & \ttt{5.5}{39} \\
AD1 B01    & \ttt{1.5}{3} & \ttt{1.5}{3} \hfill ($96\%$) & \ttt{4.9}{1} \hfill ($3\%$) & \ttt{1.3}{42} & \ttt{5.5}{40} \\
AD1 BK     & \ttt{5.0}{3} & \ttt{4.9}{3} \hfill ($99\%$) & \ttt{1.6}{1} \hfill ($0\%$) & \ttt{4.0}{43} & \ttt{1.7}{42} \\
AD1 BS     & \ttt{3.7}{3} & \ttt{3.6}{3} \hfill ($98\%$) & \ttt{5.8}{1} \hfill ($1\%$) & \ttt{8.6}{41} & \ttt{3.6}{40} \\
 \enddata
\label{tab:results_sf100}
\tablecomments{The same as in Tab.~\ref{tab:results_sfc}, but for burst SF which started $100\myr$ ago and lasted for $100\myr$. }
\end{deluxetable*}

\newpage
\begin{deluxetable*}{lccccc}
    \tablewidth{\textwidth}
    \tablecaption{General parameters of the simulational results (burst SFR $1\gyr$ ago)}
    \tablehead{Model\tablenotemark{a} & \multicolumn{3}{c}{Number per
        MWEG\tablenotemark{a}} & \multicolumn{2}{c}{$\lxmax
        [\ergs]$\tablenotemark{b}} \\
& \nulx & \nbhulx & \nnsulx & BHULX & NSULX }
    \startdata
 & \multicolumn{5}{c}{$\zsun$}\\
AD0 BN     & \ttt{8.3}{1} & \ttt{8.5}{-3} \hfill ($0\%$) & \ttt{8.3}{1} \hfill ($99\%$) & \ttt{2.0}{42} & \ttt{2.2}{42} \\
AD0 B01    & \ttt{3.9}{2} & \ttt{3.6}{-1} \hfill ($0\%$) & \ttt{3.9}{2} \hfill ($99\%$) & \ttt{2.0}{43} & \ttt{2.2}{43} \\
AD0 BK     & \ttt{6.0}{1} & \ttt{3.5}{-4} \hfill ($0\%$) & \ttt{6.0}{1} \hfill ($99\%$) & \ttt{6.2}{44} & \ttt{6.8}{44} \\
AD0 BS     & \ttt{5.5}{1} & \ttt{1.5}{-3} \hfill ($0\%$) & \ttt{5.5}{1} \hfill ($99\%$) & \ttt{1.3}{43} & \ttt{1.4}{43} \\
AD1 BN     & \ttt{3.8}{1} & \ttt{8.7}{-5} \hfill ($0\%$) & \ttt{3.8}{1} \hfill ($99\%$) & \ttt{3.7}{39} & \ttt{3.9}{39} \\
AD1 B01    & \ttt{5.2}{2} & \ttt{1.2}{-1} \hfill ($0\%$) & \ttt{5.2}{2} \hfill ($99\%$) & \ttt{3.7}{40} & \ttt{3.9}{40} \\
AD1 BK     & \ttt{1.8}{1} & \ttt{2.8}{-7} \hfill ($0\%$) & \ttt{1.8}{1} \hfill ($99\%$) & \ttt{1.1}{42} & \ttt{1.2}{42} \\
AD1 BS     & \ttt{8.0}{1} & \ttt{1.3}{-5} \hfill ($0\%$) & \ttt{8.0}{1} \hfill ($99\%$) & \ttt{2.4}{40} & \ttt{2.6}{40} \\
 & \multicolumn{5}{c}{$\zsun/10$}\\
AD0 BN     & \ttt{6.8}{2} & \ttt{4.1}{0} \hfill ($0\%$) & \ttt{6.8}{2} \hfill ($99\%$) & \ttt{4.8}{42} & \ttt{1.4}{43} \\
AD0 B01    & \ttt{8.5}{2} & \ttt{1.2}{1} \hfill ($1\%$) & \ttt{8.4}{2} \hfill ($98\%$) & \ttt{4.8}{43} & \ttt{1.4}{44} \\
AD0 BK     & \ttt{5.7}{2} & \ttt{3.9}{0} \hfill ($0\%$) & \ttt{5.7}{2} \hfill ($99\%$) & \ttt{1.5}{45} & \ttt{4.3}{45} \\
AD0 BS     & \ttt{2.8}{2} & \ttt{4.2}{0} \hfill ($1\%$) & \ttt{2.8}{2} \hfill ($98\%$) & \ttt{3.2}{43} & \ttt{9.2}{43} \\
AD1 BN     & \ttt{1.2}{2} & \ttt{4.0}{0} \hfill ($3\%$) & \ttt{1.2}{2} \hfill ($96\%$) & \ttt{9.0}{39} & \ttt{3.5}{39} \\
AD1 B01    & \ttt{9.9}{2} & \ttt{5.0}{0} \hfill ($0\%$) & \ttt{9.8}{2} \hfill ($99\%$) & \ttt{9.0}{40} & \ttt{3.5}{40} \\
AD1 BK     & \ttt{1.1}{2} & \ttt{4.0}{0} \hfill ($3\%$) & \ttt{1.1}{2} \hfill ($96\%$) & \ttt{2.8}{42} & \ttt{1.1}{42} \\
AD1 BS     & \ttt{2.9}{2} & \ttt{3.8}{0} \hfill ($1\%$) & \ttt{2.9}{2} \hfill ($98\%$) & \ttt{5.9}{40} & \ttt{2.3}{40} \\

 & \multicolumn{5}{c}{$\zsun/100$}\\
AD0 BN     & \ttt{2.2}{2} & \ttt{5.6}{-2} \hfill ($0\%$) & \ttt{2.2}{2} \hfill ($99\%$) & \ttt{3.4}{42} & \ttt{7.3}{42} \\
AD0 B01    & \ttt{1.7}{2} & \ttt{1.6}{0} \hfill ($0\%$) & \ttt{1.7}{2} \hfill ($99\%$) & \ttt{3.4}{43} & \ttt{7.3}{43} \\
AD0 BK     & \ttt{1.6}{2} & \ttt{1.3}{-2} \hfill ($0\%$) & \ttt{1.6}{2} \hfill ($99\%$) & \ttt{1.0}{45} & \ttt{2.2}{45} \\
AD0 BS     & \ttt{7.2}{1} & \ttt{1.1}{-2} \hfill ($0\%$) & \ttt{7.2}{1} \hfill ($99\%$) & \ttt{2.2}{43} & \ttt{4.8}{43} \\
AD1 BN     & \ttt{3.1}{1} & \ttt{0.0}{0} \hfill ($0\%$) & \ttt{3.1}{1} \hfill ($100\%$) & \ttt{0.0}{0} & \ttt{2.8}{39} \\
AD1 B01    & \ttt{2.2}{2} & \ttt{0.0}{0} \hfill ($0\%$) & \ttt{2.2}{2} \hfill ($100\%$) & \ttt{0.0}{0} & \ttt{2.8}{40} \\
AD1 BK     & \ttt{2.9}{1} & \ttt{0.0}{0} \hfill ($0\%$) & \ttt{2.9}{1} \hfill ($100\%$) & \ttt{0.0}{0} & \ttt{8.6}{41} \\
AD1 BS     & \ttt{6.3}{1} & \ttt{0.0}{0} \hfill ($0\%$) & \ttt{6.3}{1} \hfill ($100\%$) & \ttt{0.0}{0} & \ttt{1.8}{40} \\

 \enddata
\label{tab:results_sf1000}
\tablecomments{The same as in Tab.~\ref{tab:results_sfc}, but for burst SF which started $1\gyr$ ago and lasted for $100\myr$. }
\end{deluxetable*}

\begin{deluxetable*}{lccccc}
    \tablewidth{\textwidth}
    \tablecaption{General parameters of the simulational results (burst SFR $5\gyr$ ago)}
    \tablehead{Model\tablenotemark{a} & \multicolumn{3}{c}{Number per
        MWEG\tablenotemark{a}} & \multicolumn{2}{c}{$\lxmax
        [\ergs]$\tablenotemark{b}} \\
& \nulx & \nbhulx & \nnsulx & BHULX & NSULX }
    \startdata
 & \multicolumn{5}{c}{$\zsun$}\\
AD0 BN     & \ttt{7.3}{0} & \ttt{0.0}{0} \hfill ($0\%$) & \ttt{7.3}{0} \hfill ($100\%$) & \ttt{0.0}{0} & \ttt{7.2}{43} \\
AD0 B01    & \ttt{6.6}{1} & \ttt{0.0}{0} \hfill ($0\%$) & \ttt{6.6}{1} \hfill ($100\%$) & \ttt{0.0}{0} & \ttt{7.2}{44} \\
AD0 BK     & \ttt{4.9}{0} & \ttt{0.0}{0} \hfill ($0\%$) & \ttt{4.9}{0} \hfill ($100\%$) & \ttt{0.0}{0} & \ttt{2.2}{46} \\
AD0 BS     & \ttt{3.5}{0} & \ttt{0.0}{0} \hfill ($0\%$) & \ttt{3.5}{0} \hfill ($100\%$) & \ttt{0.0}{0} & \ttt{4.7}{44} \\
AD1 BN     & \ttt{7.2}{-1} & \ttt{0.0}{0} \hfill ($0\%$) & \ttt{7.2}{-1} \hfill ($100\%$) & \ttt{0.0}{0} & \ttt{2.9}{39} \\
AD1 B01    & \ttt{5.0}{1} & \ttt{0.0}{0} \hfill ($0\%$) & \ttt{5.0}{1} \hfill ($100\%$) & \ttt{0.0}{0} & \ttt{2.9}{40} \\
AD1 BK     & \ttt{9.5}{-1} & \ttt{0.0}{0} \hfill ($0\%$) & \ttt{9.5}{-1} \hfill ($100\%$) & \ttt{0.0}{0} & \ttt{8.9}{41} \\
AD1 BS     & \ttt{2.0}{0} & \ttt{0.0}{0} \hfill ($0\%$) & \ttt{2.0}{0} \hfill ($100\%$) & \ttt{0.0}{0} & \ttt{1.9}{40} \\
 & \multicolumn{5}{c}{$\zsun/10$}\\
AD0 BN     & \ttt{6.7}{0} & \ttt{0.0}{0} \hfill ($0\%$) & \ttt{6.7}{0} \hfill ($100\%$) & \ttt{0.0}{0} & \ttt{9.4}{42} \\
AD0 B01    & \ttt{6.5}{1} & \ttt{7.2}{-1} \hfill ($1\%$) & \ttt{6.4}{1} \hfill ($98\%$) & \ttt{6.1}{39} & \ttt{9.4}{43} \\
AD0 BK     & \ttt{4.9}{0} & \ttt{0.0}{0} \hfill ($0\%$) & \ttt{4.9}{0} \hfill ($100\%$) & \ttt{0.0}{0} & \ttt{2.9}{45} \\
AD0 BS     & \ttt{4.4}{0} & \ttt{0.0}{0} \hfill ($0\%$) & \ttt{4.4}{0} \hfill ($100\%$) & \ttt{0.0}{0} & \ttt{6.2}{43} \\
AD1 BN     & \ttt{1.4}{0} & \ttt{0.0}{0} \hfill ($0\%$) & \ttt{1.4}{0} \hfill ($100\%$) & \ttt{0.0}{0} & \ttt{3.0}{39} \\
AD1 B01    & \ttt{5.1}{1} & \ttt{0.0}{0} \hfill ($0\%$) & \ttt{5.1}{1} \hfill ($100\%$) & \ttt{0.0}{0} & \ttt{3.0}{40} \\
AD1 BK     & \ttt{7.0}{-1} & \ttt{0.0}{0} \hfill ($0\%$) & \ttt{7.0}{-1} \hfill ($99\%$) & \ttt{0.0}{0} & \ttt{9.2}{41} \\
AD1 BS     & \ttt{3.2}{0} & \ttt{0.0}{0} \hfill ($0\%$) & \ttt{3.2}{0} \hfill ($100\%$) & \ttt{0.0}{0} & \ttt{2.0}{40} \\

 & \multicolumn{5}{c}{$\zsun/100$}\\
AD0 BN     & \ttt{1.9}{0} & \ttt{0.0}{0} \hfill ($0\%$) & \ttt{1.9}{0} \hfill ($100\%$) & \ttt{0.0}{0} & \ttt{1.2}{43} \\
AD0 B01    & \ttt{1.9}{1} & \ttt{2.0}{-2} \hfill ($0\%$) & \ttt{1.9}{1} \hfill ($99\%$) & \ttt{6.8}{39} & \ttt{1.2}{44} \\
AD0 BK     & \ttt{1.4}{0} & \ttt{0.0}{0} \hfill ($0\%$) & \ttt{1.4}{0} \hfill ($100\%$) & \ttt{0.0}{0} & \ttt{3.7}{45} \\
AD0 BS     & \ttt{2.0}{0} & \ttt{0.0}{0} \hfill ($0\%$) & \ttt{2.0}{0} \hfill ($100\%$) & \ttt{0.0}{0} & \ttt{7.9}{43} \\
AD1 BN     & \ttt{3.0}{-4} & \ttt{0.0}{0} \hfill ($0\%$) & \ttt{3.0}{-4} \hfill ($100\%$) & \ttt{0.0}{0} & \ttt{3.0}{39} \\
AD1 B01    & \ttt{4.9}{0} & \ttt{1.2}{-1} \hfill ($2\%$) & \ttt{4.8}{0} \hfill ($97\%$) & \ttt{6.5}{39} & \ttt{3.0}{40} \\
AD1 BK     & \ttt{1.0}{-1} & \ttt{0.0}{0} \hfill ($0\%$) & \ttt{1.0}{-1} \hfill ($100\%$) & \ttt{0.0}{0} & \ttt{9.2}{41} \\
AD1 BS     & \ttt{7.9}{-1} & \ttt{0.0}{0} \hfill ($0\%$) & \ttt{7.9}{-1} \hfill ($100\%$) & \ttt{0.0}{0} & \ttt{2.0}{40} \\
 \enddata
\label{tab:results_sf5000}
\tablecomments{The same as in Tab.~\ref{tab:results_sfc}, but for burst SF which started $5\gyr$ ago and lasted for $100\myr$. }
\end{deluxetable*}

\begin{deluxetable*}{llllll}
    \tablewidth{\textwidth}
    \tablecaption{General parameters of the simulational results (burst SFR $10\gyr$ ago)}
    \tablehead{Model\tablenotemark{a} & \multicolumn{3}{c}{Number per
        MWEG\tablenotemark{a}} & \multicolumn{2}{c}{$\lxmax
        [\ergs]$\tablenotemark{b}} \\
& \nulx & \nbhulx & \nnsulx & BHULX & NSULX }
    \startdata
 & \multicolumn{5}{c}{$\zsun$}\\
AD0 BN     & \ttt{1.2}{0} & \ttt{0.0}{0} \hfill ($0\%$) & \ttt{1.2}{0} \hfill ($100\%$) & \ttt{0.0}{0} & \ttt{8.7}{42} \\
AD0 B01    & \ttt{8.1}{0} & \ttt{0.0}{0} \hfill ($0\%$) & \ttt{8.1}{0} \hfill ($100\%$) & \ttt{0.0}{0} & \ttt{8.7}{43} \\
AD0 BK     & \ttt{8.3}{-1} & \ttt{0.0}{0} \hfill ($0\%$) & \ttt{8.3}{-1} \hfill ($100\%$) & \ttt{0.0}{0} & \ttt{2.7}{45} \\
AD0 BS     & \ttt{7.4}{-1} & \ttt{0.0}{0} \hfill ($0\%$) & \ttt{7.4}{-1} \hfill ($100\%$) & \ttt{0.0}{0} & \ttt{5.7}{43} \\
AD1 BN     & \ttt{2.9}{-2} & \ttt{0.0}{0} \hfill ($0\%$) & \ttt{2.9}{-2} \hfill ($100\%$) & \ttt{0.0}{0} & \ttt{3.0}{39} \\
AD1 B01    & \ttt{2.6}{0} & \ttt{0.0}{0} \hfill ($0\%$) & \ttt{2.6}{0} \hfill ($100\%$) & \ttt{0.0}{0} & \ttt{3.0}{40} \\
AD1 BK     & \ttt{1.1}{-1} & \ttt{0.0}{0} \hfill ($0\%$) & \ttt{1.1}{-1} \hfill ($100\%$) & \ttt{0.0}{0} & \ttt{9.2}{41} \\
AD1 BS     & \ttt{3.6}{-1} & \ttt{0.0}{0} \hfill ($0\%$) & \ttt{3.6}{-1} \hfill ($100\%$) & \ttt{0.0}{0} & \ttt{2.0}{40} \\
 & \multicolumn{5}{c}{$\zsun/10$}\\
AD0 BN     & \ttt{1.5}{0} & \ttt{0.0}{0} \hfill ($0\%$) & \ttt{1.5}{0} \hfill ($100\%$) & \ttt{0.0}{0} & \ttt{5.4}{42} \\
AD0 B01    & \ttt{1.2}{1} & \ttt{0.0}{0} \hfill ($0\%$) & \ttt{1.2}{1} \hfill ($100\%$) & \ttt{0.0}{0} & \ttt{5.4}{43} \\
AD0 BK     & \ttt{1.1}{0} & \ttt{0.0}{0} \hfill ($0\%$) & \ttt{1.1}{0} \hfill ($100\%$) & \ttt{0.0}{0} & \ttt{1.7}{45} \\
AD0 BS     & \ttt{9.4}{-1} & \ttt{0.0}{0} \hfill ($0\%$) & \ttt{9.4}{-1} \hfill ($100\%$) & \ttt{0.0}{0} & \ttt{3.6}{43} \\
AD1 BN     & \ttt{2.1}{-1} & \ttt{0.0}{0} \hfill ($0\%$) & \ttt{2.1}{-1} \hfill ($100\%$) & \ttt{0.0}{0} & \ttt{2.8}{39} \\
AD1 B01    & \ttt{7.3}{0} & \ttt{0.0}{0} \hfill ($0\%$) & \ttt{7.3}{0} \hfill ($100\%$) & \ttt{0.0}{0} & \ttt{2.8}{40} \\
AD1 BK     & \ttt{7.5}{-2} & \ttt{0.0}{0} \hfill ($0\%$) & \ttt{7.5}{-2} \hfill ($100\%$) & \ttt{0.0}{0} & \ttt{8.6}{41} \\
AD1 BS     & \ttt{2.9}{-1} & \ttt{0.0}{0} \hfill ($0\%$) & \ttt{2.9}{-1} \hfill ($100\%$) & \ttt{0.0}{0} & \ttt{1.8}{40} \\

 & \multicolumn{5}{c}{$\zsun/100$}\\
AD0 BN     & \ttt{6.2}{-1} & \ttt{0.0}{0} \hfill ($0\%$) & \ttt{6.2}{-1} \hfill ($100\%$) & \ttt{0.0}{0} & \ttt{5.1}{42} \\
AD0 B01    & \ttt{5.9}{0} & \ttt{0.0}{0} \hfill ($0\%$) & \ttt{5.9}{0} \hfill ($100\%$) & \ttt{0.0}{0} & \ttt{5.1}{43} \\
AD0 BK     & \ttt{4.4}{-1} & \ttt{0.0}{0} \hfill ($0\%$) & \ttt{4.4}{-1} \hfill ($100\%$) & \ttt{0.0}{0} & \ttt{1.6}{45} \\
AD0 BS     & \ttt{3.8}{-1} & \ttt{0.0}{0} \hfill ($0\%$) & \ttt{3.8}{-1} \hfill ($100\%$) & \ttt{0.0}{0} & \ttt{3.4}{43} \\
AD1 BN     & \ttt{1.5}{-4} & \ttt{0.0}{0} \hfill ($0\%$) & \ttt{1.5}{-4} \hfill ($100\%$) & \ttt{0.0}{0} & \ttt{2.8}{39} \\
AD1 B01    & \ttt{2.9}{0} & \ttt{0.0}{0} \hfill ($0\%$) & \ttt{2.9}{0} \hfill ($100\%$) & \ttt{0.0}{0} & \ttt{2.8}{40} \\
AD1 BK     & \ttt{5.0}{-7} & \ttt{0.0}{0} \hfill ($0\%$) & \ttt{5.0}{-7} \hfill ($100\%$) & \ttt{0.0}{0} & \ttt{8.6}{41} \\
AD1 BS     & \ttt{2.3}{-5} & \ttt{0.0}{0} \hfill ($0\%$) & \ttt{2.3}{-5} \hfill ($100\%$) & \ttt{0.0}{0} & \ttt{1.8}{40} \\
 \enddata
\label{tab:results_sf10000}
\tablecomments{The same as in Tab.~\ref{tab:results_sfc}, but for burst SF which
started $10\gyr$ ago and lasted for $100\myr$. A total absence of BHULXs may be
observed.}
\end{deluxetable*}

\newpage

\newpage

\newpage
\begin{deluxetable*}{l||cccc|c|cc||ccc|ccc|ccc}
    \tablewidth{\textwidth}
    \tablecaption{Companions (constant SFR)}
    \tablehead{Model & MS   & HG   & RG   & CHeB & HeMS & HeWD & HybWD& MS   & HG   & RG   & HeMS & HeHG & HeGB & HeWD & COWD & HybWD}
    \startdata
 & \multicolumn{16}{c}{$\zsun$}\\
AD0 BN     & 0.82 & 0.18 & 0.01 &      &      &      &      & 0.03 & 0.05 & 0.05 & 0.01 & 0.08 &      & 0.32 & 0.09 & 0.37\\
AD0 B01    & 0.91 & 0.07 & 0.02 &      &      &      &      & 0.02 & 0.02 & 0.01 & 0.16 &      &      & 0.38 & 0.10 & 0.32\\
AD0 BK     & 0.91 & 0.09 &      &      &      &      &      & 0.01 & 0.04 & 0.04 & 0.01 & 0.08 &      & 0.34 & 0.10 & 0.39\\
AD0 BS     & 0.92 & 0.08 &      &      &      &      &      & 0.01 & 0.06 & 0.03 & 0.03 & 0.09 &      & 0.33 & 0.09 & 0.35\\
AD1 BN     & 0.85 & 0.15 &      &      &      &      &      & 0.04 & 0.10 & 0.02 & 0.02 & 0.09 & 0.01 & 0.22 & 0.08 & 0.42\\
AD1 B01    & 0.94 & 0.05 &      &      &      &      &      & 0.05 & 0.04 & 0.01 & 0.15 & 0.01 &      & 0.33 & 0.10 & 0.32\\
AD1 BK     & 0.94 & 0.06 &      &      &      &      &      & 0.13 & 0.10 & 0.08 & 0.04 & 0.04 &      & 0.17 & 0.07 & 0.37\\
AD1 BS     & 0.94 & 0.06 &      &      &      &      &      & 0.09 & 0.16 & 0.05 & 0.02 & 0.08 &      & 0.23 & 0.08 & 0.30\\
 & \multicolumn{16}{c}{$\zsun/10$}\\
AD0 BN     & 0.68 & 0.30 &      & 0.01 &      &      &      & 0.06 & 0.47 & 0.02 &      & 0.01 &      & 0.21 & 0.07 & 0.17\\
AD0 B01    & 0.87 & 0.10 &      &      &      &      &      & 0.02 & 0.03 &      & 0.03 &      &      & 0.47 & 0.15 & 0.29\\
AD0 BK     & 0.74 & 0.24 & 0.01 & 0.01 &      &      &      & 0.03 & 0.51 & 0.01 &      &      &      & 0.21 & 0.07 & 0.16\\
AD0 BS     & 0.79 & 0.20 & 0.01 & 0.01 &      &      &      & 0.03 & 0.49 & 0.01 &      & 0.01 &      & 0.22 & 0.07 & 0.17\\
AD1 BN     & 0.69 & 0.28 &      & 0.02 &      &      &      & 0.10 & 0.39 & 0.02 &      & 0.02 &      & 0.19 & 0.08 & 0.21\\
AD1 B01    & 0.91 & 0.08 &      &      &      &      &      & 0.06 & 0.04 & 0.01 & 0.03 &      &      & 0.42 & 0.15 & 0.30\\
AD1 BK     & 0.76 & 0.21 & 0.01 & 0.01 &      &      &      & 0.11 & 0.54 & 0.05 &      &      &      & 0.13 & 0.05 & 0.13\\
AD1 BS     & 0.81 & 0.17 & 0.01 & 0.01 &      &      &      & 0.08 & 0.52 & 0.03 &      & 0.01 &      & 0.16 & 0.06 & 0.15\\
 & \multicolumn{16}{c}{$\zsun/100$}\\
AD0 BN     & 0.77 & 0.09 &      & 0.13 & 0.01 &      &      & 0.02 & 0.48 & 0.04 &      & 0.06 &      & 0.25 & 0.01 & 0.13\\
AD0 B01    & 0.89 & 0.04 &      & 0.06 &      &      &      & 0.03 & 0.03 & 0.03 & 0.07 &      &      & 0.56 & 0.03 & 0.25\\
AD0 BK     & 0.83 & 0.05 &      & 0.12 &      &      &      & 0.01 & 0.48 & 0.04 &      & 0.04 &      & 0.27 & 0.01 & 0.14\\
AD0 BS     & 0.87 & 0.05 &      & 0.08 &      &      &      & 0.02 & 0.29 & 0.12 &      & 0.04 &      & 0.35 & 0.02 & 0.18\\
AD1 BN     & 0.76 & 0.10 &      & 0.13 & 0.01 &      &      & 0.04 & 0.45 & 0.05 &      & 0.13 &      & 0.14 & 0.01 & 0.18\\
AD1 B01    & 0.91 & 0.04 &      & 0.05 &      &      &      & 0.13 & 0.05 & 0.06 & 0.07 &      &      & 0.33 & 0.03 & 0.31\\
AD1 BK     & 0.84 & 0.06 &      & 0.10 &      &      &      & 0.12 & 0.47 & 0.20 &      & 0.03 &      & 0.07 & 0.01 & 0.10\\
AD1 BS     & 0.89 & 0.05 &      & 0.06 &      &      &      & 0.10 & 0.36 & 0.27 &      & 0.03 &      & 0.11 & 0.01 & 0.12
 \enddata
 \tablecomments{Companions (donors) in ULXs formed in models with constant SFR. Fractions for different companion types are provided separately for BHULXs and NSULXs. The age of the population is $10\gyr$.}
\label{tab:comps_sfc}
\end{deluxetable*}

\begin{deluxetable*}{l||cccc|c|cc||ccc|ccc|ccc}
    \tablewidth{\textwidth}
    \tablecaption{Companions (burst SFR 100\myr\ ago)}
    \tablehead{Model & MS   & HG   & RG   & CHeB & HeMS & HeWD & HybWD& MS   & HG   & RG   & HeMS & HeHG & HeGB & HeWD & COWD & HybWD}
    \startdata

 & \multicolumn{16}{c}{$\zsun$}\\
AD0 BN     & 0.89 & 0.11 &      &      &      &      &      & 0.42 &      &      & 0.02 & 0.49 & 0.01 &      &      & 0.06\\
AD0 B01    & 0.96 & 0.03 &      &      &      &      &      & 0.01 &      &      & 0.98 & 0.01 &      &      &      &     \\
AD0 BK     & 0.97 & 0.03 &      &      &      &      &      & 0.29 &      &      & 0.04 & 0.57 &      &      &      & 0.10\\
AD0 BS     & 0.97 & 0.03 &      &      &      &      &      & 0.07 &      &      & 0.67 & 0.22 &      &      &      & 0.03\\
AD1 BN     & 0.90 & 0.10 &      &      &      &      &      & 0.28 &      &      &      & 0.59 & 0.12 &      &      &     \\
AD1 B01    & 0.97 & 0.03 &      &      &      &      &      & 0.14 &      &      & 0.85 & 0.01 &      &      &      &     \\
AD1 BK     & 0.98 & 0.02 &      &      &      &      &      & 0.89 &      &      &      & 0.10 &      &      &      &     \\
AD1 BS     & 0.97 & 0.03 &      &      &      &      &      & 0.73 &      &      & 0.11 & 0.15 & 0.01 &      &      &     \\
 & \multicolumn{16}{c}{$\zsun/10$}\\
AD0 BN     & 0.86 & 0.11 &      & 0.02 & 0.01 &      &      & 0.98 &      &      & 0.01 & 0.01 &      &      &      &     \\
AD0 B01    & 0.97 & 0.03 &      &      &      &      &      & 0.97 &      &      & 0.02 & 0.01 &      &      &      &     \\
AD0 BK     & 0.94 & 0.04 &      & 0.01 &      &      &      & 0.98 &      &      & 0.02 & 0.01 &      &      &      &     \\
AD0 BS     & 0.95 & 0.03 &      & 0.01 &      &      &      & 0.96 &      &      & 0.03 & 0.01 &      &      &      &     \\
AD1 BN     & 0.85 & 0.12 &      & 0.03 & 0.01 &      &      & 0.94 &      &      & 0.01 & 0.04 & 0.02 &      &      &     \\
AD1 B01    & 0.97 & 0.03 &      & 0.01 &      &      &      & 0.99 &      &      & 0.01 &      &      &      &      &     \\
AD1 BK     & 0.94 & 0.03 &      & 0.02 &      &      &      & 1.00 &      &      &      &      &      &      &      &     \\
AD1 BS     & 0.95 & 0.04 &      & 0.01 &      &      &      & 0.99 &      &      &      &      &      &      &      &     \\
 & \multicolumn{16}{c}{$\zsun/100$}\\
AD0 BN     & 0.81 & 0.05 &      & 0.13 & 0.01 &      &      & 0.97 &      &      &      & 0.02 &      &      &      &     \\
AD0 B01    & 0.91 & 0.02 &      & 0.06 &      &      &      & 0.88 &      &      & 0.10 & 0.01 &      &      &      &     \\
AD0 BK     & 0.86 & 0.01 &      & 0.12 &      &      &      & 0.98 &      &      & 0.01 & 0.01 &      &      &      & 0.01\\
AD0 BS     & 0.90 & 0.02 &      & 0.08 &      &      &      & 0.85 &      &      & 0.12 & 0.01 &      &      &      & 0.01\\
AD1 BN     & 0.79 & 0.06 &      & 0.14 & 0.01 &      &      & 0.60 &      &      &      & 0.27 & 0.13 &      &      &     \\
AD1 B01    & 0.92 & 0.02 &      & 0.05 &      &      &      & 0.98 &      &      & 0.01 & 0.01 &      &      &      &     \\
AD1 BK     & 0.87 & 0.02 &      & 0.11 &      &      &      & 1.00 &      &      &      &      &      &      &      &     \\
AD1 BS     & 0.92 & 0.02 &      & 0.06 &      &      &      & 0.98 &      &      &      & 0.02 &      &      &      &     

 \enddata
 \tablecomments{Companions (donors) in ULXs formed in models with burst SF. Start of the burst was $100\myr$ ago and its duration was $100\myr$.}
\label{tab:comps_sf100}
\end{deluxetable*}

\begin{deluxetable*}{l||cccc|c|cc||ccc|ccc|ccc}
    \tablewidth{\textwidth}
    \tablecaption{Companions (burst SFR 1\gyr\ ago)}
    \tablehead{Model & MS   & HG   & RG   & CHeB & HeMS & HeWD & HybWD& MS   & HG   & RG   & HeMS & HeHG & HeGB & HeWD & COWD & HybWD}
    \startdata

 & \multicolumn{16}{c}{$\zsun$}\\
AD0 BN     &      & 1.00 &      &      &      &      &      & 0.02 & 0.07 &      &      &      &      & 0.49 & 0.09 & 0.33\\
AD0 B01    & 0.28 & 0.43 & 0.29 &      &      &      &      & 0.07 & 0.03 &      &      &      &      & 0.12 & 0.03 & 0.74\\
AD0 BK     &      & 1.00 &      &      &      &      &      & 0.01 & 0.07 &      &      &      &      & 0.49 & 0.09 & 0.33\\
AD0 BS     &      & 1.00 &      &      &      &      &      & 0.02 & 0.12 &      &      &      &      & 0.46 & 0.08 & 0.31\\
AD1 BN     &      &      & 1.00 &      &      &      &      & 0.01 & 0.31 &      & 0.01 &      &      & 0.35 & 0.07 & 0.25\\
AD1 B01    & 0.51 & 0.01 & 0.47 &      & 0.01 &      &      & 0.15 & 0.17 & 0.02 &      &      &      & 0.09 & 0.02 & 0.54\\
AD1 BK     &      &      & 1.00 &      &      &      &      & 0.03 & 0.41 &      & 0.01 &      &      & 0.29 & 0.06 & 0.20\\
AD1 BS     &      &      & 1.00 &      &      &      &      & 0.04 & 0.47 &      &      &      &      & 0.26 & 0.05 & 0.17\\
 & \multicolumn{16}{c}{$\zsun/10$}\\
AD0 BN     & 0.40 & 0.37 & 0.18 &      &      &      &      & 0.01 & 0.74 &      &      &      &      & 0.17 & 0.03 & 0.05\\
AD0 B01    & 0.32 & 0.44 & 0.23 &      &      &      &      & 0.03 & 0.08 &      &      &      &      & 0.32 & 0.05 & 0.51\\
AD0 BK     & 0.41 & 0.34 & 0.19 &      &      &      &      &      & 0.78 &      &      &      &      & 0.15 & 0.03 & 0.05\\
AD0 BS     & 0.40 & 0.37 & 0.18 &      &      &      &      & 0.01 & 0.58 &      &      &      &      & 0.28 & 0.05 & 0.08\\
AD1 BN     & 0.60 & 0.10 & 0.24 &      &      &      &      & 0.04 & 0.47 &      &      &      &      & 0.33 & 0.06 & 0.11\\
AD1 B01    & 0.64 & 0.22 & 0.13 &      &      &      &      & 0.08 & 0.11 &      &      &      &      & 0.28 & 0.05 & 0.47\\
AD1 BK     & 0.60 & 0.10 & 0.24 &      &      &      &      & 0.03 & 0.73 & 0.01 &      &      &      & 0.15 & 0.03 & 0.05\\
AD1 BS     & 0.58 & 0.12 & 0.25 &      &      &      &      & 0.02 & 0.63 & 0.01 &      &      &      & 0.23 & 0.04 & 0.07\\
 & \multicolumn{16}{c}{$\zsun/100$}\\
AD0 BN     &      & 0.87 & 0.03 & 0.10 &      &      &      &      & 0.79 &      &      &      &      & 0.17 & 0.01 & 0.03\\
AD0 B01    & 0.12 & 0.47 & 0.38 &      &      &      & 0.03 & 0.01 & 0.10 & 0.01 &      &      &      & 0.36 & 0.01 & 0.51\\
AD0 BK     &      & 1.00 &      &      &      &      &      &      & 0.79 &      &      &      &      & 0.17 & 0.01 & 0.03\\
AD0 BS     &      & 0.90 & 0.02 & 0.07 &      &      &      &      & 0.60 &      &      &      &      & 0.32 & 0.02 & 0.05\\
AD1 BN     &      &      &      &      &      &      &      &      & 0.46 & 0.02 &      &      &      & 0.43 & 0.03 & 0.07\\
AD1 B01    &      &      &      &      &      &      &      & 0.19 & 0.08 & 0.02 &      &      &      & 0.29 & 0.01 & 0.42\\
AD1 BK     &      &      &      &      &      &      &      & 0.12 & 0.57 & 0.10 &      &      &      & 0.17 & 0.01 & 0.03\\
AD1 BS     &      &      &      &      &      &      &      & 0.05 & 0.46 & 0.12 &      &      &      & 0.30 & 0.02 & 0.04
 \enddata
 \tablecomments{Companions (donors) in ULXs formed in models with burst SF. Start of the burst was 1\gyr\ ago and its duration was 100\myr.}
\label{tab:comps_sf1000}
\end{deluxetable*}

\begin{deluxetable*}{l||cccc|c|cc||ccc|ccc|ccc}
    \tablewidth{\textwidth}
    \tablecaption{Companions (burst SFR 5\gyr\ ago)}
    \tablehead{Model & MS   & HG   & RG   & CHeB & HeMS & HeWD & HybWD& MS   & HG   & RG   & HeMS & HeHG & HeGB & HeWD & COWD & HybWD}
    \startdata
 & \multicolumn{16}{c}{$\zsun$}\\
AD0 BN     &      &      &      &      &      &      &      &      &      & 0.37 &      &      &      & 0.63 &      &     \\
AD0 B01    &      &      &      &      &      &      &      &      &      & 0.01 &      &      &      & 0.75 & 0.24 &     \\
AD0 BK     &      &      &      &      &      &      &      &      &      & 0.32 &      &      &      & 0.68 &      &     \\
AD0 BS     &      &      &      &      &      &      &      &      &      & 0.19 &      &      &      & 0.81 &      &     \\
AD1 BN     &      &      &      &      &      &      &      &      &      & 0.31 &      &      &      & 0.69 &      &     \\
AD1 B01    &      &      &      &      &      &      &      &      &      & 0.02 &      &      &      & 0.63 & 0.35 &     \\
AD1 BK     &      &      &      &      &      &      &      &      &      & 0.78 &      &      &      & 0.22 &      &     \\
AD1 BS     &      &      &      &      &      &      &      &      &      & 0.61 &      &      &      & 0.39 &      &     \\
 & \multicolumn{16}{c}{$\zsun/10$}\\
AD0 BN     &      &      &      &      &      &      &      &      &      & 0.16 &      &      &      & 0.56 & 0.28 &     \\
AD0 B01    &      &      &      &      &      &      &      & 0.01 &      & 0.02 &      &      &      & 0.70 & 0.27 &     \\
AD0 BK     &      &      &      &      &      &      &      &      &      & 0.17 &      &      &      & 0.55 & 0.28 &     \\
AD0 BS     &      &      &      &      &      &      &      &      &      & 0.21 &      &      &      & 0.53 & 0.26 &     \\
AD1 BN     &      &      &      &      &      &      &      &      &      & 0.02 &      &      &      & 0.43 & 0.54 &     \\
AD1 B01    &      &      &      &      &      &      &      & 0.01 &      & 0.04 &      &      &      & 0.54 & 0.41 &     \\
AD1 BK     &      &      &      &      &      &      &      &      &      & 0.25 &      &      &      & 0.34 & 0.41 &     \\
AD1 BS     &      &      &      &      &      &      &      &      &      & 0.38 &      &      &      & 0.28 & 0.34 &     \\
 & \multicolumn{16}{c}{$\zsun/100$}\\
AD0 BN     &      &      &      &      &      &      &      & 0.10 &      & 0.25 &      &      &      & 0.66 &      &     \\
AD0 B01    &      &      &      &      &      & 1.00 &      & 0.02 & 0.03 & 0.17 &      &      &      & 0.77 &      &     \\
AD0 BK     &      &      &      &      &      &      &      & 0.04 &      & 0.30 &      &      &      & 0.66 &      &     \\
AD0 BS     &      &      &      &      &      &      &      & 0.02 &      & 0.60 &      &      &      & 0.38 &      &     \\
AD1 BN     &      &      &      &      &      &      &      &      &      & 1.00 &      &      &      &      &      &     \\
AD1 B01    &      &      &      &      &      & 0.17 &      & 0.09 & 0.10 & 0.68 &      &      &      & 0.08 &      &     \\
AD1 BK     &      &      &      &      &      &      &      & 0.38 &      & 0.62 &      &      &      &      &      &     \\
AD1 BS     &      &      &      &      &      &      &      & 0.03 &      & 0.97 &      &      &      &      &      &     

 \enddata
 \tablecomments{Companions (donors) in ULXs formed in models with burst SF. Start of the burst was 5\gyr\ ago and its duration was 100\myr.}
\label{tab:comps_sf5000}
\end{deluxetable*}

\begin{deluxetable*}{l||cccc|c|cc||ccc|ccc|ccc}
    \tablewidth{\textwidth}
    \tablecaption{Companions (burst SFR 10\gyr\ ago)}
    \tablehead{Model & MS   & HG   & RG   & CHeB & HeMS & HeWD & HybWD& MS   & HG   & RG   & HeMS & HeHG & HeGB & HeWD & COWD & HybWD}
    \startdata
 & \multicolumn{16}{c}{$\zsun$}\\
AD0 BN     &      &      &      &      &      &      &      &      &      & 0.40 &      &      &      & 0.60 &      &     \\
AD0 B01    &      &      &      &      &      &      &      &      &      & 0.08 &      &      &      & 0.92 &      &     \\
AD0 BK     &      &      &      &      &      &      &      &      &      & 0.38 &      &      &      & 0.62 &      &     \\
AD0 BS     &      &      &      &      &      &      &      &      &      & 0.40 &      &      &      & 0.60 &      &     \\
AD1 BN     &      &      &      &      &      &      &      &      &      & 0.19 &      &      &      & 0.81 &      &     \\
AD1 B01    &      &      &      &      &      &      &      &      &      & 0.31 &      &      &      & 0.69 &      &     \\
AD1 BK     &      &      &      &      &      &      &      &      &      & 0.92 &      &      &      & 0.08 &      &     \\
AD1 BS     &      &      &      &      &      &      &      &      &      & 0.91 &      &      &      & 0.09 &      &     \\
 & \multicolumn{16}{c}{$\zsun/10$}\\
AD0 BN     &      &      &      &      &      &      &      &      &      &      &      &      &      & 0.83 & 0.17 &     \\
AD0 B01    &      &      &      &      &      &      &      &      &      & 0.07 &      &      &      & 0.73 & 0.20 &     \\
AD0 BK     &      &      &      &      &      &      &      &      &      &      &      &      &      & 0.83 & 0.17 &     \\
AD0 BS     &      &      &      &      &      &      &      &      &      &      &      &      &      & 0.83 & 0.17 &     \\
AD1 BN     &      &      &      &      &      &      &      &      &      &      &      &      &      & 0.89 & 0.11 &     \\
AD1 B01    &      &      &      &      &      &      &      &      &      & 0.11 &      &      &      & 0.56 & 0.34 &     \\
AD1 BK     &      &      &      &      &      &      &      &      &      &      &      &      &      & 0.90 & 0.10 &     \\
AD1 BS     &      &      &      &      &      &      &      &      &      &      &      &      &      & 0.89 & 0.11 &     \\
 & \multicolumn{16}{c}{$\zsun/100$}\\
AD0 BN     &      &      &      &      &      &      &      &      &      &      &      &      &      & 1.00 &      &     \\
AD0 B01    &      &      &      &      &      &      &      & 0.05 & 0.07 & 0.12 &      &      &      & 0.54 &      &     \\
AD0 BK     &      &      &      &      &      &      &      &      &      &      &      &      &      & 1.00 &      &     \\
AD0 BS     &      &      &      &      &      &      &      &      &      &      &      &      &      & 1.00 &      &     \\
AD1 BN     &      &      &      &      &      &      &      &      &      & 1.00 &      &      &      &      &      &     \\
AD1 B01    &      &      &      &      &      &      &      & 0.11 & 0.13 & 0.27 &      &      &      & 0.01 &      &     \\
AD1 BK     &      &      &      &      &      &      &      &      &      & 1.00 &      &      &      &      &      &     \\
AD1 BS     &      &      &      &      &      &      &      &      &      & 1.00 &      &      &      &      &      &    
 \enddata
 \tablecomments{Companions (donors) in ULXs formed in models with burst SF.
 Start of the burst was 10\gyr\ ago and its duration was 100\myr. A total
 absence of BHULXs may be observed.}
\label{tab:comps_sf10000}
\end{deluxetable*}

\end{document}